\documentclass[11pt,letter]{article}
\usepackage[top=1in,bottom=1in,left=1in,right=1in]{geometry}
\pdfoutput = 1
\usepackage[noEucal]{main}
\usepackage{aas_macros}
\usepackage[utf8]{inputenc}
\makeatletter \g@addto@macro\@floatboxreset\centering \makeatother
\usepackage{graphicx}
\usepackage[scr=boondoxo]{mathalfa}
\usepackage[bbgreekl]{mathbbol}
\DeclareSymbolFontAlphabet{\mathbbm}{bbold}
\usepackage[font={small,it}]{caption}
\usepackage{mathrsfs}

\newcommand{\sbra}[1]{  [\, #1 \, | } 
\newcommand{\sket}[1]{  | \, #1  \,] } 
\newcommand{\savg}[1]{ [\, #1 \, ]}
\def\bh{{\bar h}}

\begin{document}
\begin{titlepage}
\unitlength = 1mm
\ \\
\vskip 3cm
\begin{center}

{\LARGE{\textsc{Bulk Wilson Loops from $\CI$}}}

\vspace{0.8cm}
James Drummond\,$^a$, Bruno Gim\'enez Umbert\,$^{a,b}$ and Prahar Mitra\,$^a$

\vspace{1cm}

{\it $^a$\,\,Physics \& Astronomy and Mathematical Sciences,\\
University of Southampton, Southampton, SO17 1BJ UK}\\
{\it $^b$\,\,Department of Physics \& Astronomy,\\ University of Waterloo, Waterloo, ON N2L 3G1, Canada}

\vspace{0.8cm}

\begin{abstract}
A Carrollian description of four-dimensional Yang-Mills theory on partially complexified future null infinity was recently proposed by Opreij, Skinner and Wang in \cite{Opreij:2026bdx}. In this paper, we give a new form of the Carrollian Yang-Mills kinetic action on real future null infinity, and recover the logarithmic two-point function found by \cite{Opreij:2026bdx}. We then extend the Carrollian description of Yang-Mills beyond scattering amplitudes by studying the expectation value of a null polygonal Wilson loop in Minkowski spacetime. We propose a boundary operator given by the trace of a cyclic product of holomorphic frames on the good cuts associated with the polygon's vertices. At order $g^2$, we show analytically that the boundary calculation reproduces the rational bulk integrand obtained from a self-dual Lagrangian insertion.
\end{abstract}

\vspace{1.0cm}
\end{center}
\end{titlepage}
\pagestyle{empty}
\pagestyle{plain}
\pagenumbering{arabic}

\tableofcontents

\section{Introduction}
\label{sec:intro}

Finding a holographic description of quantum gravity in asymptotically flat spacetimes remains an important open problem. The natural observables in this context are scattering amplitudes of massless particles, which propagate out to null infinity, and this motivates looking for a description of bulk dynamics in terms of degrees of freedom living on the asymptotic null boundary. Two closely related approaches have been developed in recent years along these lines. In celestial holography \cite{Strominger:2017zoo, Pasterski:2021rjz, Raclariu:2021zjz}, scattering amplitudes in 4-dimensional flat spacetime are formulated as correlators of conformal primary operators in a putative CFT$_2$ on the celestial sphere, via a Mellin transform in the energies of the external particles. Bulk asymptotic symmetries then become infinite-dimensional symmetry algebras acting on these correlators, and soft theorems appear as Ward identities. In parallel, in Carrollian holography \cite{Donnay:2022aba, Bagchi:2022emh, Donnay:2022wvx, Nguyen:2023vfz, Nguyen:2023miw} the boundary theory lives on null infinity itself and has conformal Carrollian symmetry. Here, the bulk energy is traded for the retarded time by a Fourier transform rather than a Mellin transform, and Carrollian correlators retain the dependence on position along the null generators. These two descriptions are related by integral transforms \cite{Donnay:2022wvx}.

Most of the work in both directions has focused on reformulating scattering amplitudes. However, if a boundary theory is to be holographic in the spirit of AdS/CFT, it should also reproduce bulk observables beyond scattering amplitudes, including observables that are not built from asymptotic states at all. In this paper, we make progress in this direction by constructing in Yang-Mills theory a boundary dual to a null polygonal Wilson loop in Minkowski space and showing that its rational integrand at $\CO(g^2)$ agrees with the chiral bulk integrand.

Our starting point is the recent proposal of \cite{Opreij:2026bdx}, which constructs an action for Yang-Mills on a partial complexification of future null infinity $\CI^+$, denoted $\CI_\mcc^+$. The action combines an electric-branch Carrollian kinetic term with non-local interactions of MHV type, and was shown to reproduce the tree-level MHV and NMHV amplitudes. The construction is closely related to the twistor action for Yang-Mills \cite{Boels:2006ir}, in which non-local interactions on twistor space encode local spacetime dynamics. As emphasised in \cite{Opreij:2026bdx}, this is not a dual in the strict AdS/CFT sense: the fundamental fields are the asymptotic values of the bulk gauge field rather than independent boundary degrees of freedom. The authors argue that this may be an unavoidable feature of asymptotically flat spacetimes, where radiation reaches the boundary instead of being reflected by it, so a flat space holographic dictionary does not need to take the same form as in AdS/CFT.

We first give a formulation of the kinetic action on $\CI^+$. In the construction of \cite{Opreij:2026bdx}, the authors considered a partial complexification of future null infinity $\CI_\mcc^+$ in which the null time $u$ on $\CI^+$ is allowed to be complex. The local kinetic action on $\CI_\mcc^+$ leads to a logarithmic two-point function for the gauge field, which was independently derived in \cite{Donnay:2022wvx}. This logarithmic structure plays a crucial role in reproducing the bulk MHV and NMHV amplitudes from the boundary action. We show here it is possible to reproduce this logarithmic structure from a non-local action on $\CI^+$, with the non-locality implemented by a Hilbert transform in $u$ \cite{King_2009}. Such non-localities, while surprising in traditional holographic settings such as AdS/CFT, are expected in Carrollian holographic duals \cite{Cotler:2025npu}. This allows us to reformulate the Carrollian action of \cite{Opreij:2026bdx} on $\CI^+$ instead of $\CI^+_\mcc$. 

The observable we consider in this work is the expectation value of a null polygonal Wilson loop. This is a natural choice for various reasons. First, null Wilson loops are dual to MHV amplitudes in the planar limit of $\CN=4$ super Yang-Mills \cite{Alday:2007hr,Drummond:2007aua,Brandhuber:2007yx,Drummond:2007cf,Bern:2008ap,Drummond:2008aq} and, in their supersymmetric extension, to all planar amplitudes \cite{Mason:2010yk,Caron-Huot:2010ryg}. They are thus closely tied to the amplitudes the action was built to reproduce, while being genuinely non-local bulk observables. Second, they are naturally adapted to the geometry of $\CI^+$. Every bulk point $x$ determines a good cut $C_x$, which is the intersection of its future light cone with $\CI^+$, i.e. a copy of the celestial sphere. Two bulk points are null separated precisely when their cuts touch at a single point, which lies on the generator of $\CI^+$ labelled by the null direction joining them. A null polygon with vertices $x_i$ is therefore encoded in a closed chain of cuts $C_{x_i}$, with consecutive cuts meeting at the points where the null edges reach $\CI^+$ (see Figure \ref{Fig-WLgeom}). Moreover, exactly this structure appears in twistor space. In \cite{Mason:2010yk}, a null polygon is represented by a cyclic chain of twistor lines: the vertices correspond to lines, the edges to the twistors at which consecutive lines meet, and the Wilson loop is the trace of a product of holomorphic frames along these lines. The action of \cite{Opreij:2026bdx} contains the same ingredients, namely holomorphic frames $U_x$ on the cuts $C_x \cong \mathbb{CP}^1$, which play the role of Wilson lines for a Riemann surface rather than for a curve.

Inspired by this picture, we propose a boundary operator dual to the bulk null Wilson loop: the trace of a product of holomorphic frames, one on each cut $C_{x_i}$, glued together at the points where consecutive cuts meet. We compute its expectation value in the Carrollian theory at the first non-trivial order in the Yang-Mills coupling constant $g$, namely $\CO(g^2)$. At this order the Wilson loop is determined by single gluon exchange, so its expectation value and the integrand defined via the Lagrangian insertion in Section \ref{sec:bulk_calc} coincide in both pure Yang-Mills and in $\CN=4$ super Yang-Mills. This allows a direct comparison with results obtained in the latter theory.

To start with, we first present the computation of the null polygonal Wilson loop in the bulk. Using a scaling argument, we relate the $\CO(g^2)$ term in the Wilson loop expectation value to the integrated connected two-point function of the Lagrangian operator and the Wilson loop itself. The integral here is over the insertion point $x_0$ of the Lagrangian. The integrand, namely the two-point function, is UV finite and is only defined up to total derivatives in $x_0$, which do not affect the integrated result. Chiral formulations, like the twistor description, naturally produce parity non-invariant integrands. To compare our result with those, 
we use the chiral form of the Yang-Mills Lagrangian, which is twice its self-dual part. This differs from the real Lagrangian by a $\t$-term with $\t=\frac{8\pi^2 i}{g^2}$. This chiral form of the Yang-Mills Lagrangian gives the self-dual bulk integrand.

On the boundary, the integral over good cuts in the interaction term plays the role of the integral over $x_0$. Our main result is that the boundary computation reproduces this self-dual one-loop integrand, which can equivalently be written as the sum over chiral pentagons \cite{Arkani-Hamed:2010pyv}
familiar from the twistor-space formulation of the Wilson loop \cite{Mason:2010yk}. We verify this agreement at the level of rational integrands analytically. The precise matching requires a cancellation of several spurious poles. This shows that, at least at $\CO(g^2)$, the Carrollian formulation of \cite{Opreij:2026bdx} computes a non-local bulk observable that does not involve asymptotic states, extending its reach beyond the scattering amplitudes for which it was constructed.

Our results suggest several directions for future work. An obvious next step is to consider higher order corrections to the Wilson loop expectation value. At $\CO(g^4)$, the Wilson loop receives contributions from the self-interactions of the Yang-Mills gauge field and from loop corrections. This would extend the present comparison beyond single-gluon exchange. Another interesting extension involves the connected correlators of multiple null polygonal Wilson loops \cite{Drummond:2025ulh}. It is also interesting to consider the supersymmetric extension of the Carrollian action, specifically to $\CN=4$ SYM. A boundary dual of the super Wilson loop would give, at the level of the integrand, the full planar S-matrix of $\CN=4$ SYM normalised by the MHV tree amplitude \cite{Mason:2010yk, Caron-Huot:2010ryg}.\footnote{At the integrated level, regularisation introduces additional subtleties. In particular, the spacetime super Wilson loop can acquire anomalous contributions that break chiral supersymmetry \cite{Belitsky:2011zm}. At one loop, subtracting these contributions restores agreement with reduced non-MHV amplitudes \cite{Belitsky:2012nu}.} 

Moreover, correlation functions of certain local gauge-invariant operators in $\CN=4$ SYM, normalised by their tree-level values, are related to null Wilson loops in the limit when consecutive points become light-like separated \cite{Alday:2010zy}. This correspondence was established in twistor space at the level of the integrand in \cite{Adamo:2011pv}. It would be interesting to reproduce this relation from null infinity and investigate whether the boundary description can be extended to local bulk correlators at generic separations.

The paper is organised as follows. In Section \ref{sec:scri_geometry}, we review the geometry of $\CI^+$, including flat null and homogeneous coordinates and the good cuts associated with bulk points. In Section \ref{sec:Action}, we review the Carrollian action of \cite{Opreij:2026bdx}, presenting the kinetic term without complexifying $\CI^+$. In Section \ref{sec:WL}, we define the null Wilson loop and its proposed boundary dual, and compute both at $\CO(g^2)$. In particular, in Section \ref{sec:bulk_calc} we derive the bulk integrand in its self-dual form; in Section \ref{sec:boundary_calc} we carry out the boundary computation; in Section \ref{app:analyticmatch}, we show that the self-dual bulk integrand    and the boundary integrand agree as rational functions. Appendix \ref{app:2pt_gauge} shows that the non-local Carrollian action reproduces the logarithmic two-point function and Appendix \ref{app:action_normalisation} confirms the normalisation of the Carrollian action by matching the boundary two-point function and three-point MHV amplitude to the corresponding bulk results.

\section{The Geometry of \texorpdfstring{$\CI^+$}{scri-plus}}
\label{sec:scri_geometry}

\subsection{Flat Null Coordinates}
\label{sec:flat_null}

We work in flat null coordinates $(u,r,z,\bz)$, which are related to Cartesian coordinates $x^\mu$ via \cite{Dumitrescu:2015fej, He:2020ifr}\footnote{Note that our definition of $q^\mu(z,\bz)$ differs from the standard one used in Celestial holography literature. More precisely, $q_{\text{here}}^\mu(z,\bz) = q^\mu_{\text{Celestial}}(\bz,z)$. We use the opposite convention here to match the one used in \cite{Opreij:2026bdx}. Also, our $u$ differs from the one in \cite{Dumitrescu:2015fej, He:2020ifr} by a factor of 2, $u_\text{there}=2u_\text{here}$.}
\begin{equation}
\begin{split}
\label{qn_def}
x^\mu = r q^\mu(z,\bz) + u n^\mu \,,  \quad q^\mu(z,\bz) \equiv \left( \frac{1 + z\bz}{2} , \frac{z+\bz}{2} , \frac{\bz-z}{2i}, \frac{1-z\bz}{2} \right) \,, \quad n^\mu \equiv  (1,0,0,-1)\,, 
\end{split}
\end{equation}
where $u \in \mrr$, $r \in \mrr^*$ and $z \in \mcc$. The metric of four-dimensional Minkowski spacetime $M_4$ is
\begin{equation}
\dt s^2 = \eta_{\mu\nu} \dt x^\mu \dt x^\nu =  - 2 \dt u \dt r + r^2 \dt z \dt \bz \,,
\end{equation}
which follows from the fact that $q(z,\bz)^2=n^2=0$ and $n \cdot q(z,\bz) = -1$. In these coordinates, $\CI^+$ is located at $r=\infty$ keeping $(u,z,\bz)$ fixed. Coordinates on $\CI^+$ are $(u,z,\bz)$.

\paragraph{Lorentz Transformations:} These coordinates make manifest the isomorphism between the four dimensional Lorentz group and the global Euclidean conformal group on the two-dimensional celestial sphere, $\mathsf{SO}^\uparrow(1,3) \cong \mathsf{PSL}(2,\mcc)=\mathsf{SL}(2,\mcc)/\mzz_2$. Parameterising a Lorentz matrix by
\begin{equation}
\begin{split}
\label{Lorentz_Matrix}
\L^\mu{}_\nu = - \frac{1}{2} P^\a{}_\b ( P^* )^\da{}_\db (\s^\mu)_{\a\da} ({\bar \s}_\nu)^{\db\b}  \,, \qquad P^\a{}_\b = \begin{bmatrix}
 d & c \\ b & a 
\end{bmatrix} \,, \qquad P \sim - P \,, \qquad \det P = +1\,,
\end{split}
\end{equation}
with $(\s^\mu)_{\a\da} = ( I_2,\vec{\s})$, $({\bar \s}^\mu)^{\da\a} = (I_2,-\vec{\s})$, we find that the coordinates on $\CI^+$ transform as \cite{He:2020ifr}
\begin{equation}
\begin{split}
\label{Mobius_Transformation}
(u,z,\bz) \to ( u | \tau'(z) | , \tau(z), {\bar \tau}(\bz) ) \,, \qquad \tau(z) \equiv \frac{az+b}{cz+d} \,, \qquad {\bar \tau}(\bz) = \tau(z)^*\,.
\end{split}
\end{equation}
More details on these coordinates can be found in \cite{He:2020ifr}.

In a gravitational theory, the Lorentz symmetry \eqref{Mobius_Transformation} is enhanced locally to a Virasoro symmetry (a.k.a. \emph{superrotations}) which acts just as above but where $\tau(z)$ is an arbitrary (locally) holomorphic function \cite{Banks:2003vp, Barnich:2010ojg}. The Ward identity for this Virasoro symmetry is the Cachazo-Strominger subleading soft graviton theorem \cite{Cachazo:2014fwa,Kapec:2014opa} and the leading logarithmic soft graviton theorem \cite{Sahoo:2018lxl, Choi:2024ygx, Choi:2024ajz}.

\paragraph{Translations:} The coordinates $(u,z,\bz)$ also transform nicely under translations. In Cartesian coordinates, translations act as $x^\mu \to x^\mu + a^\mu$. Moving to flat null coordinates and taking the large $r$ limit to $\CI^+$, we find
\begin{equation}
\begin{split}
\label{translations}
(u,z,\bz) \to ( u + f(z,\bz) , z,\bz  )\,, \qquad f(z,\bz) \equiv - a \cdot q(z,\bz) \,. 
\end{split}
\end{equation}
As with Lorentz symmetry, the translational symmetry \eqref{translations} is enhanced in a gravitational theory to \emph{supertranslations} which act just as above, but where $f(z,\bz)$ is an arbitrary function \cite{Bondi:1962px, Sachs:1962wk} (not necessarily holomorphic). The Ward identity for supertranslations is Weinberg's leading soft graviton theorem \cite{Weinberg:1965aa, He:2014laa}.

\paragraph{Carroll primaries:} The transformations \eqref{Mobius_Transformation} (denoted by \rm{L}) and \eqref{translations} (denoted by \rm{T}) on $\CI^+$ are together referred to as \emph{Carroll transformations}. A Carroll primary operator with weight $(h,\bh)$ transforms under Carroll transformations as \cite{Bagchi:2016bcd, Donnay:2022aba}
\begin{equation}
\begin{split}
\label{carroll_primary_transformations}
\CO (u,z,\bz) ~\xrightarrow{\rm{L}}~ \CO'(u,z,\bz)  &= \tau'(z)^{h} {\bar \tau}'(\bz)^{\bh} \CO ( u|\tau'(z)| ,\tau(z),{\bar \tau}(\bz) )  \,, \\
\CO (u,z,\bz) ~\xrightarrow{\rm{T}}~ \CO'(u,z,\bz) &= \CO ( u + f(z,\bz),z,\bz) \,.
\end{split}
\end{equation}
It is easy to see that if $\CO$ is a Carroll primary with weights $(h,\bh)$, then $\p_u^m \CO$ is also a Carroll primary of weight $(h+\frac{m}{2},\bh+\frac{m}{2})$.

\subsection{Good Cuts}
\label{sec:good_cuts}

A transverse cut $C_\a$ of $\CI^+$ is defined in general by a hypersurface equation $u=\a(z,\bz)$. Coordinates on $C_\a$ are $(z,\bz)$. For generic functions $\a$, the cuts $C_\a$ do not transform covariantly under the Carroll transformations. A \emph{good cut} $C_x$ of $\CI^+$ is parameterised by a point $x^\mu$ in the interior of Minkowski spacetime and is defined by the equation
\begin{equation}
\begin{split}
\label{good_cut_eq}
u = - x \cdot q(z,\bz)\,.
\end{split}
\end{equation}
\begin{figure}
\includegraphics{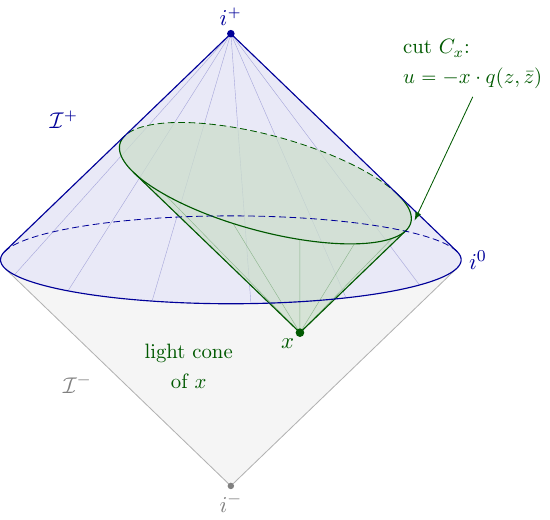}
\caption{A good cut $C_x$ of $\CI^+$ is the intersection of the closure of the future lightcone of $x$ with $\CI^+$. We also refer to it as the celestial sphere of $x$.}
\label{fig:good_cut}
\end{figure}
$C_x$ is the intersection of the closure of the future lightcone of $x$ with $\CI^+$ (Figure \ref{fig:good_cut}),
\begin{equation}
\begin{split}
C_x = \overline{L_x^+} \cap \CI^+ \,. 
\end{split}
\end{equation}
To see this, we note that the future light-cone $L_x^+$ of $x^\mu$ consists of all points that are of the form\footnote{Points along $n^\mu$ are obtained by taking $\l \to 0$ and $|w| \to \infty$ keeping $\l|w|^2=s$ fixed.}
\begin{equation}
\begin{split}
x^\mu  + \l q^\mu(w,\bw)  \,, \qquad \l \in \mrr_+ \,, \quad w \in \mcc \,. 
\end{split}
\end{equation}
The coordinates of this point in flat null coordinates are
\begin{equation}
\begin{split}
u &= - x \cdot q(w,\bw)  + O(\l^{-1}) \,, \\
r &= \l - x \cdot n + O(\l^{-1}) \,, \\
z &= w + O(\l^{-1})\,, \\
\bz &= \bw + O(\l^{-1})\,. 
\end{split}
\end{equation}
The intersection of the closure of the future light cone with $\CI^+$ is the space of all points at $\l=\infty$. We see from the above that all these points have $r=\infty$ and therefore live on $\CI^+$. We also see that the coordinate $u$ on this boundary is restricted to $- x \cdot q(w,\bw)$ which is precisely \eqref{good_cut_eq}.

Poincar\'e transformations map the good cut $C_x$ to the cut $C_{\L x+a}$, with the point $(z,\bz) \in C_x$ mapped to $(\tau(z),{\bar \tau}(\bz)) \in C_{\L x+a}$. This follows from the identity
\begin{equation}
\begin{split}
\L^\mu{}_\nu q^\nu(z,\bz) = \frac{1}{|\tau'(z)|} q^\mu(\tau(z),{\bar \tau}(\bz))  \,. 
\end{split}
\end{equation}

A Carroll primary $\CO$ induced onto a good cut $C_x$ is denoted by $[\CO]_x$. Explicitly,
\begin{equation}
\begin{split}
\label{good_cut_op}
[\CO]_x(z,\bz) = \CO ( - x \cdot q(z,\bz),z,\bz) \,. 
\end{split}
\end{equation}
Under Carrollian transformations,
\begin{equation}
\begin{split}
\label{goodcut_op_LT}
[\CO]_x(z,\bz) ~\xrightarrow{\text{L}+\text{T}} ~  \tau'(z)^{h} {\bar \tau}'(\bz)^{\bh} [\CO]_{\L x+a} ( \tau(z) , {\bar \tau}(\bz) ) \,. 
\end{split}
\end{equation}

\subsection{Homogeneous Coordinates}
\label{sec:homogeneous_coord}

We will find it convenient to describe $\CI^+$ in homogeneous coordinates $(\msu,\l_\a,{\bar \l}_\da)$ (with $\msu \in \mrr$) instead of the coordinates $(u,z,\bz)$. These will be used often in Sections \ref{sec:boundary_calc} and \ref{app:analyticmatch}. 

To describe these, we first introduce 2-component Weyl spinors $\l^\a,{\bar \l}^\da \neq 0$ with ${\bar \l}^\da = (\l^\a)^*$ that transform in the fundamental and conjugate-fundamental representations of $\mathsf{SL}(2,\mcc)$,
\begin{equation}
\begin{split}
\l^\a \to P^\a{}_\b \l^\b \,, \qquad {\bar \l}^\da \to ( P^* )^\da{}_\db {\bar \l}^\db \,. 
\end{split}
\end{equation}
Spinor indices are raised and lowered with the invariant tensors $\e_{\a\b}$ and $\e_{\da\db}=(\e_{\a\b})^*$ of $\mathsf{SL}(2,\mcc)$, which we normalise by $\e_{12}=\e^{12}=+1$, e.g.
\begin{equation}
\l^\a = \e^{\a \b}\l_{\b}\,, \qquad \l_\a = \l^{\b} \e_{\b \a}\,. 
\end{equation}
We will often use spinor-helicity notation and denote
\begin{equation}
\begin{split}
\label{spinor_defs}
\l^\a \equiv \bra{\l}^\a  \,, \qquad 
\l_\a \equiv \ket{\l}_\a \,, \qquad 
{\bar \l}_\da  \equiv \sbra{{\bar \l}}_\da \,, \qquad
{\bar \l}^\da \equiv \sket{{\bar\l}}^\da \,. 
\end{split}
\end{equation}
Spinor contractions are then
\begin{equation}
\begin{split}
\avg{\l_1\l_2} \equiv \bra{\l_1}^\a \ket{\l_2}_\a \,, \qquad \savg{{\bar \l}_1{\bar \l}_2} \equiv \sbra{{\bar \l}_1}_\da \sket{{\bar \l}_2}^\da  \,. 
\end{split}
\end{equation}
For contractions with a four-vector $q^\mu$, we define
\begin{equation}
\begin{split}
\bra{\l_1} \,q\, \sket{{\bar \l}_2} \equiv - q_\mu \bra{\l_1}^\a (\s^\mu)_{\a\da} \sket{{\bar \l}_2}^\da\,, \qquad \sbra{{\bar \l}_1} \,q\, \ket{\l_2} \equiv - q_\mu \sbra{{\bar \l}_1}_\da ({\bar \s}^\mu)^{\da\a} \ket{\l_2}_\a \,. 
\end{split}
\end{equation}
Homogeneous coordinates $(\msu,\l_\a,{\bar \l}_\da)$ on $\CI^+$ are defined modulo the scaling
\begin{equation}
\begin{split}
\label{scaling_freedom}
(\msu,\l_\a,{\bar \l}_\da) \sim (|\zeta|^2 \msu, \zeta \l_\a ,{\bar \zeta} {\bar \l}_\da )\,, \qquad \zeta \in \mcc^*\,.
\end{split}
\end{equation}
In the coordinate patch where $\l_2,{\bar \l}_2 \neq 0$, the homogeneous coordinates are related to the flat null coordinates via
\begin{equation}
\begin{split}
(u,z,\bz) = \left( \frac{2\msu}{\bra{\l} \,n\, \sket{\bar \l}} , - \l_1 / \l_2 , - {\bar \l}_1/{\bar \l}_2 \right) \,. 
\end{split}
\end{equation}
Note that this relationship is invariant under \eqref{scaling_freedom}. In this coordinate patch where $\l_2,{\bar \l}_2 \neq 0$, we can use the rescaling freedom \eqref{scaling_freedom} to set
\begin{equation}
\begin{split}
\label{spinor_coord_patch}
\l^\a = ( 1 , z ) \,, \qquad \l_\a = ( - z , 1 ) \,, \qquad {\bar \l}^\da = ( 1 , \bz ) \,, \qquad {\bar \l}_\da = ( - \bz , 1 )  \quad \implies \quad  \msu = u \,. 
\end{split}
\end{equation}
With this normalisation, we note the properties
\begin{equation}
\begin{split}
\l_\a {\bar \l}_\da &= - q_\mu(z,\bz) ( \s^\mu)_{\a\da} \,, \qquad \avg{ \l \dt \l } = - \dt z  \,, \qquad \savg{ {\bar \l} \dt {\bar \l} } = \dt \bz \,,
\end{split}
\end{equation}
and
\begin{equation}
\begin{split}
\avg{\l_1 \l_2} = z_{12} \equiv z_1 - z_2 \,, \qquad \savg{ {\bar \l}_1 {\bar \l}_2 } = - \bz_{12} \,.
\end{split}
\end{equation}
Finally, we note that the good cut equation \eqref{good_cut_eq} in homogeneous coordinates is
\begin{equation}
\begin{split}
\msu = \frac{1}{2} \bra{\l}\,x\,\sket{{\bar \l}} \,. 
\end{split}
\end{equation}

\section{The Carrollian Action}
\label{sec:Action}

Recently, a formulation of Yang-Mills theory at null infinity has been proposed and has been used to compute certain tree-level Yang-Mills scattering amplitudes \cite{Opreij:2026bdx}. The dynamical data in the formulation correspond to the limiting boundary behaviour of the Yang-Mills fields as one approaches the future null boundary $\CI^+$. We refer the reader to \cite{Opreij:2026bdx} for the details. Here we provide a very brief review of the construction.

\subsection{Gauge Field on \texorpdfstring{$\CI^+$}{scri-plus}}
\label{sec:gauge_field_scri}

The Yang-Mills gauge field is a Lie-algebra-valued one-form
\begin{equation}
\begin{split}
\CA = \CA_\mu \dt x^\mu = \CA_\mu^I T^I \dt x^\mu  \,, 
\end{split}
\end{equation}
where $T^I$ are the generators of the Lie algebra $\mfg$ associated to the Lie group $G$. These satisfy $[T^I,T^J]=f^{IJK}T^K$ with $f^{IJK} \in \mrr$.

Near $\CI^+$ (i.e., at large $r$), the flat null components of the gauge field fall off as (see \cite{He:2015zea})
\begin{equation}
\begin{split}
\CA_u = O(1/r) \,, \qquad \CA_r = O(1/r^2) \,, \qquad \CA_z = O(1) \,. 
\end{split}
\end{equation}
The gauge field on $\CI^+$ is then
\begin{equation}
A \equiv \lim_{r \to \infty} \CA = a + {\bar a} \,, \qquad a \equiv A_{z}(u,z,\bz) \dt z \,, \qquad  {\bar a} \equiv A_\bz(u,z,\bz) \dt \bz  \,.
\end{equation}
The boundary degrees of freedom on $\CI^+$ are thus the one-forms $a$ and ${\bar a}$. The positive-frequency parts of $a$ and ${\bar a}$ annihilate positive- and negative-helicity gluons, respectively, as shown in Appendix \ref{app:kinetic_normalization}. Note that this is the opposite of the convention that is typically used in Celestial holography literature, e.g. in \cite{Strominger:2017zoo}. However, it is the convention employed by \cite{Opreij:2026bdx} so we shall stick to it in this paper.

Under Carroll transformations, the gauge field on $\CI^+$ transforms as
\begin{equation}
\begin{split}
\label{gauge_LT}
A_{z}(u,z,\bz) &\xrightarrow{\rm L} \tau'(z) A_{z}(u|\tau'(z)|,\tau(z),{\bar \tau}(\bz)) \,, \qquad A_{\bz}(u,z,\bz) \xrightarrow{\rm L} {\bar \tau}'(\bz) A_{\bz} (u |\tau'(z)|,\tau(z),{\bar \tau}(\bz))\,, \\
A_{z}(u,z,\bz) &\xrightarrow{\rm T} A_{z}(u+f(z,\bz),z,\bz) \,, \qquad \qquad \quad~ A_{\bz}(u,z,\bz) \xrightarrow{\rm T} A_{\bz} (u+f(z,\bz),z,\bz) \,.
\end{split}
\end{equation}
Comparing these with \eqref{carroll_primary_transformations}, we see that $A_z$ and $A_\bz$ are Carroll primaries with weights $(1,0)$ and $(0,1)$ respectively. The one-forms $a$ and ${\bar a}$ have weights $(0,0)$.

We normalise the gauge field so that the Yang-Mills Lagrangian form is
\begin{equation}
\begin{split}
\label{YM_Lagrangian}
\CL_{bulk} = - \frac{1}{4} \Tr [ \CF^{\mu\nu} \CF_{\mu\nu} ] \,, \qquad \CF_{\mu\nu} = \p_\mu \CA_\nu - \p_\nu \CA_\mu + g [ \CA_\mu , \CA_\nu ] \,,
\end{split}
\end{equation}
where the Lie algebra trace here is normalised so that $\Tr [ T^I T^J ] = \d^{IJ}$, and $g$ is the Yang-Mills coupling constant. The theory is invariant under gauge transformations of the form
\begin{equation}
\begin{split}\label{gauge_symm}
\CA \to h \CA h^{-1} + \frac{1}{g} h \dt h^{-1} \,, \qquad h : M_4 \to G \,. 
\end{split}
\end{equation}
For functions that approach the identity element ${\mathbb 1} \in G$ at infinity, \eqref{gauge_symm} represents a redundancy in our description of the theory. These redundancies are dealt with by a choice of gauge condition. Functions $h$ that approach to $\msg(z,\bz) \neq {\mathbb 1}$ on $\CI^+$ are physical symmetries of the theory known as \emph{large gauge symmetries}. Under large gauge symmetries the gauge field on $\CI^+$ transforms as
\begin{equation}
\begin{split}
\label{LGT}
A_z \to \msg A_z \msg^{-1} + \frac{1}{g} \msg \p_z \msg^{-1} \,, \qquad A_\bz \to \msg A_\bz \msg^{-1} +  \frac{1}{g} \msg \p_\bz \msg^{-1} \,.
\end{split}
\end{equation}

\subsection{Kinetic Action}

The kinetic term in the Carrollian action of \cite{Opreij:2026bdx} is quadratic in the forms $a$ and ${\bar a}$ and gives rise to the two-point function of the theory. To describe this action, \cite{Opreij:2026bdx} uses a complex extension of $\CI^+$, allowing $u$ to take on complex values. In this section, we present an alternative formulation of the action that does not require this complexification and that leads to the same two-point function, up to $i\e$ prescriptions which we deal with separately, on the real section of complexified null infinity as required for the calculations in this paper.\footnote{The overall normalisation of the two-point function in \cite{Opreij:2026bdx} seems to differ from the one here (the normalisation conventions used in that paper are not clear). We have independently confirmed our normalisation by matching the boundary two-point to the bulk two-point function in Appendix \ref{app:kinetic_normalization}.}

The kinetic action is
\begin{equation}
\begin{split}
\label{kinetic_action}
S_{kin} = - 8i \int_{\CI^+} \dt u \dt^2 z \Tr [ \p_u A_z \CL (A_\bz) ] \,, \qquad \dt^2 z = i  \dt z \w \dt \bz \,. 
\end{split}
\end{equation}
where $\CL$ denotes the Hilbert-transform \cite{King_2009},
\begin{equation}
\begin{split}
\CL (A_\bz)(u,z,\bz) \equiv \frac{1}{2\pi} \text{PV} \int_{-\infty}^\infty \dt u' \frac{A_\bz(u',z,\bz)}{u-u'} \,. 
\end{split}
\end{equation}
The singularity at $u'=u$ in the Hilbert transform is dealt with using the principal value. Another subtlety here is that the gauge field on $\CI^+$ does not die off near $\CI^+_\pm$ (i.e. as $u \to \pm\infty$) so the Hilbert transform is divergent. The boundary values of the gauge field control the infrared sector of the gauge theory. These boundary modes do not play a role in the calculations here. To deal with this, we should separate the gauge field $A_\bz$ into its radiative mode ${\hat A}_\bz$, a Goldstone mode $C_\bz$, and a soft mode $N_\bz$ \cite{He:2020ifr}. The action \eqref{kinetic_action} should then involve only the radiative gauge field. This falls off sufficiently fast near $\CI^+_\pm$ and the Hilbert transform of ${\hat A}_\bz$ is well-defined. The contribution of the Goldstone mode $C_\bz$ and the soft mode $N_\bz$ to the kinetic action should be related to the soft effective action derived in \cite{Kapec:2021eug}. We will explore this in future work. For the purposes of this work, we will ignore these subtleties.

This preserves all the Carrollian symmetries described in the previous section (including supertranslations and superrotations). Additionally, despite appearances, \eqref{kinetic_action} is symmetric in $A_z$ and $A_\bz$ (up to boundary terms when we integrate by parts in $u$) and we can equivalently write
\begin{equation}
\begin{split}
S_{kin} = - 8i\int_{\CI^+} \dt u  \dt^2 z \Tr [ \p_u A_\bz  \CL (A_z) ] \,.
\end{split}
\end{equation}
In other words, the kinetic part of the Carrollian action describes the positive- and negative-helicity gluons on the same footing. Notice that this also implies that on the real section where $(A_z^I)^* = A^I_\bz$, the kinetic part of the action is purely imaginary (up to boundary terms). A similar imaginary kinetic term appears also in the soft effective action that describes the infrared dynamics of gauge and gravity theories \cite{Kalyanapuram:2021bvf, Kapec:2021eug, He:2024ddb, Choi:2026dhz}.

With the Gaussian weight $e^{iS_{kin}}$, the action determines the symmetric logarithmic kernel. We show in Appendix \ref{app:2pt_gauge} that this kinetic action gives the 2-point function
\begin{equation}
\begin{split}
\label{A_2pt}
\avg{ A^I_z(u,z,\bz) A^J_\bw(u',w,\bw) } = - \frac{\d^{IJ}}{4\pi} \ln \frac{|u - u'|}{\ell_0}\d^\2(z-w) \,, 
\end{split}
\end{equation}
where $\ell_0$ is an infrared length scale. The $\delta$-function is normalised as
\begin{equation}
\begin{split}
\int \dt^2 z \d^\2(z-w) = 1 \,.
\end{split}
\end{equation}
We show in Appendix \ref{app:kinetic_normalization} that with an appropriate $i\epsilon$ prescription \eqref{A_2pt} is precisely the boundary limit of the bulk two-point function obtained from the Yang-Mills Lagrangian.

For our calculations in Section \ref{sec:boundary_calc}, we will only need the two-point function
\begin{equation}
\begin{split}
\label{A_2pt_needed}
\avg{ \p_u^2 A^I_z(u,z,\bz) A^J_\bw(u',w,\bw) } = \frac{\d^{IJ}}{4\pi} \frac{1}{( u - u' )^2 } \d^\2(z-w)\,.
\end{split}
\end{equation}
The prescription for the cut separation is specified in Section \ref{sec:boundary_calc}.

\subsection{Interaction Action}
\label{sec:interaction_action}

The interaction term is essentially a generating function for MHV amplitudes and does not treat $a$ and ${\bar a}$ symmetrically. It is\footnote{As with the quadratic action, the overall normalisation of $S_{int}$ in \cite{Opreij:2026bdx} seems to differ from the one here. We have independently confirmed our normalisation by evaluating the three-point MHV amplitude with helicities $(--+)$ from the bulk and boundary and verifying that they match (see Appendix \ref{app:interaction_normalization}).}
\begin{equation}
\begin{split}
\label{interaction_action}
S_{int} &= \frac{1}{16\pi^2} \int_{M_4} \dt^4 x \int_{\mcc^2} \dt^2 z \dt^2 z' ( z - z' )^2 \Tr \biggl( [\p_u^2 A_z]_x U_x(z,\bz|z',\bz') [\p_u^2 A_{z'}]_x  U_x(z',\bz'|z,\bz) \biggr)\,,
\end{split}
\end{equation}
where recall that $[\CO]_x$ denotes a Carrollian operator $\CO$ evaluated on the good cut $C_x$. Explicitly,
\begin{equation}
\begin{split}
[\p_u^2 A_z]_x \equiv \p_u^2 A_z(u,z,\bz) |_{u=-x \cdot q(z,\bz)} \,. 
\end{split}
\end{equation}
The holomorphic frame $U_x(z,\bz|z',\bz')$ lives on $C_x$ and is defined by conditions
\begin{equation}
\begin{split}
\label{Ux_eq_def}
( \p_\bz + g A_\bz ) |_{C_x} U_x(z,\bz|z',\bz')  = 0 \,, \qquad U_x(z',\bz'|z',\bz')  = {\mathbb 1} \,.
\end{split}
\end{equation}
The frame defines a map $U_x(P|Q)\colon E_Q \to E_P$ between the gauge-bundle fibres at $Q=(z',\bz')$ and $P=(z,\bz)$ on $C_x$, as in \cite{Opreij:2026bdx}. It acts by left multiplication on a vector in $E_Q$.

Under Carrollian symmetries,
\begin{equation}
\begin{split}
\label{Ux_LT}
U_x(z,\bz|z',\bz') \xrightarrow{\text{L}+\text{T}} U_{\L x + a} (\tau(z),{\bar \tau}(\bz)|\tau(z'),{\bar\tau}(\bz')) \,. 
\end{split}
\end{equation}
In other words, this transforms as a good cut operator with weights $(0,0)$ in $z$ and $(0,0)$ in $z'$. Under large gauge transformations \eqref{LGT}, it transforms as
\begin{equation}
\begin{split}
\label{Ux_LGT}
U_x(z,\bz|z',\bz') \xrightarrow{\text{LGT}} \msg(z,\bz) U_x(z,\bz|z',\bz') \msg(z',\bz')^{-1} \,. 
\end{split}
\end{equation}
It can be verified that the action \eqref{interaction_action} is invariant under Poincar\'e transformations discussed in Section \ref{sec:flat_null}. For example, we can count the $z$-weights of all the operators: $\dt^2 z$ has weight $(-1,-1)$, $(z-z')^2$ has weight $(-1,0)$, $\p_u^2 A_z$ has weight $(2,1)$, $\p_u^2 A_{z'}$ has weight $(0,0)$ and $U_x$ has weight $(0,0)$. The total weight of the action is $(-1,-1)+(-1,0)+(2,1)=(0,0)$. The same counting can be done for $z'$. Additionally, under the Carrollian transformations, the good cut $C_x$ also maps to the good cut $C_{\L x+a}$. This is accounted for by the integral over all $x \in M_4$. Finally, the action is also invariant under large gauge transformations \eqref{LGT} due to the trace. The interaction action does, however, break supertranslation and superrotation symmetry.

An explicit form for $U_x$ can be obtained by solving the differential equation \eqref{Ux_eq_def}. Working perturbatively around the trivial connection, this solves to
\begin{equation}
\begin{split}
\label{holomorphic_frame}
U_x(z,\bz|z',\bz') = {\mathbb 1} + \sum_{n = 1}^{\infty}  \left( - \frac{g}{2\pi} \right)^n \int_{\mcc^n}  \frac{\dt^2 z_1 \cdots \dt^2 z_n (z-z') [ A_{\bz_1} ]_x \cdots [ A_{\bz_n} ]_x }{(z-z_1)(z_1-z_2) \cdots ( z_{n-1}-z_n)(z_n-z')} \,.
\end{split}
\end{equation}

The interaction \eqref{interaction_action} is chiral. It is quadratic in $\p_u^2 A_z$, and expanding the two holomorphic frames using \eqref{holomorphic_frame} generates arbitrarily high powers of $A_\bz$. In the construction of \cite{Opreij:2026bdx}, a single vertex gives an MHV tree amplitude, while $\mathrm{N}^k\mathrm{MHV}$ tree amplitudes are obtained by joining $k+1$ such vertices with the mixed propagator. This construction parallels the CSW rules \cite{Cachazo:2004kj}, which can be derived from spacetime \cite{Mansfield:2005yd,Gorsky:2005sf} and from twistor space \cite{Boels:2007qn}.

We can also describe the action \eqref{interaction_action} in homogeneous coordinates $(\msu,\l,{\bar \l})$ as
\begin{equation}
S_{int} = \frac{1}{16\pi^2} \int_{M_4} \dt^4 x \int_{C_x \times C_x}[{\bar \l} \dt {\bar \l}] [{\bar \l}' \dt {\bar \l}'] \avg{ \l \l' }^2 \, \Tr \biggl( \p_\msu^2 a(\l)  U_x(\l|\l') \p_\msu^2 a(\l')  U_x(\l'|\l)\biggr)\,,
\label{Sint}
\end{equation}
where
\begin{equation}
\begin{split}
\p_\msu^2 a(\l) \equiv \p_\msu^2 a  ( \msu,\l,{\bar \l}  ) \big|_{\msu=\frac{1}{2} \bra{\l} x \sket{{\bar \l}} }  \,, \qquad \p_\msu^2 a(\l') \equiv \p_\msu^2 a  ( \msu,\l',{\bar \l}'  ) \big|_{\msu=\frac{1}{2} \bra{\l'} x \sket{{\bar \l}'} } \,.
\end{split}
\end{equation}
The holomorphic frame $U_x(\l|\l')$ is now given by
\begin{equation}
\label{Ux_holomorphic}
U_x(\l|\l') = {\mathbb 1} + \sum_{n = 1}^{\infty} \left(-\frac{g}{2\pi i} \right)^n \int_{(C_x)^n}\frac{\avg{ \l \l' } ( \avg{ \l_1 \dt \l_1 } \, {\bar a}_1 ) \cdots (  \avg{ \l_n \dt \l_n } \, {\bar a}_n ) }{ \avg{ \l \l_1 } \avg{ \l_1 \l_2 } \cdots \avg{ \l_{n-1} \l_n } \avg{ \l_n \l' }}  \,,
\end{equation}
where
\begin{equation}
\begin{split}
{\bar a}_i \equiv {\bar a} ( \msu , \l_i , {\bar \l}_i )  \big|_{\msu=\frac{1}{2} \bra{\l_i} x \sket{{\bar \l_i}} }  \,.
\end{split}
\end{equation}
In Appendix \ref{app:interaction_normalization}, we verify that the interaction action correctly reproduces, including the normalisations, the tree-level three-point MHV amplitude with helicities $(--+)$.

\section{Null Wilson Loops}\label{sec:WL}

It is clear from the ingredients introduced in the action \eqref{Sint}, and also highlighted in the discussion in \cite{Opreij:2026bdx}, that this formulation of Yang-Mills is very close to the twistorial description employed in e.g. \cite{Boels:2006ir}. Here we would like to illustrate this connection further with the calculation of a different observable located entirely within the bulk Minkowski space: a piecewise light-like Wilson loop. It might be surprising that the boundary theory on $\CI^+$ can be used to compute observables which do not involve the asymptotic on-shell states, however the close connection to the twistor formulation suggests a very natural quantity which we will show matches the expected answer obtained from a bulk spacetime calculation. Here we are directly motivated by the construction of \cite{Mason:2010yk} for (supersymmetric) light-like  Wilson loops in $\CN=4$ super Yang-Mills theory.

Let us first describe the relevant bulk and boundary geometry for the loop we want to consider. In the bulk, we consider a sequence of points in Minkowski space $x_1, \cdots, x_n,x_{n+1}=x_1$ such that each point is null separated from its two neighbours $(x_{i+1} -x_i)^2 = 0$. We then consider the Wilson loop
\begin{equation}
\begin{split}
\CW[C_n] = \frac{1}{d_R} \Tr_R \CP \exp \bigg( \!\! - g \oint_{C_n} \CA \bigg) \,,
\end{split}
\end{equation}
where $C_n$ is the null loop $x_1 \to x_2 \to \cdots \to x_n \to x_1$. The path-ordering operator $\CP$ controls the ordering of the Lie algebra generators in the exponential. The trace is over a representation $R$ of the Lie group $\CG$ and $d_R$ is the dimension of the representation. We compute
\begin{equation}
\begin{split}
\CW_n = \avg{ \CW[C_n]  } = \frac{1}{Z_{bulk}} \int \frac{[ \dt \CA ]}{\text{vol}(\CG)} e^{ i S_{bulk} } \CW[C_n] \,, 
\end{split}
\end{equation}
where
\begin{equation}
\begin{split}
S_{bulk} = \int_{M_4} \dt^4 x \CL_{bulk} (x) \,, 
\end{split}
\end{equation}
$\CL_{bulk}$ is the bulk Yang-Mills Lagrangian \eqref{YM_Lagrangian}. This is evaluated to $\CO(g^2)$ in Section \ref{sec:bulk_calc}.

\begin{figure}
\includegraphics{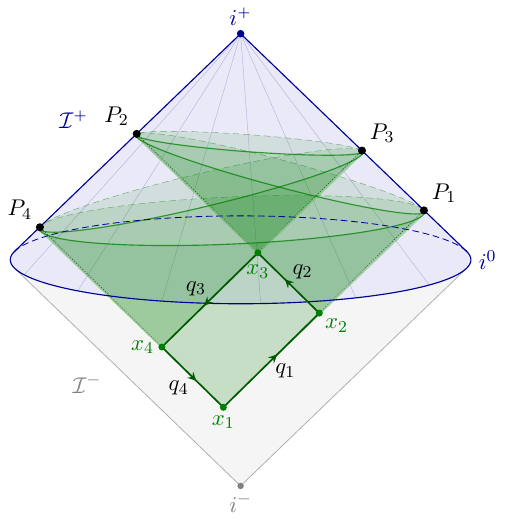}
\caption{The good cuts $C_{x_i}$ and $C_{x_{i+1}}$ meet at the point $P_i$ where the future continuation of the null edge $(x_i,x_{i+1})$ reaches $\CI^+$.}
\label{Fig-WLgeom}
\end{figure}

To describe the boundary dual of a bulk null Wilson loop, we define
\begin{equation}
\begin{split}
q_i^\mu \equiv x^\mu_{i+1} - x^\mu_i \,.
\end{split}
\end{equation}
These ``momenta'' satisfy
\begin{equation}
\begin{split}
q_i^2 = 0 \,, \qquad \sum_i q_i^\mu = 0 \,. 
\end{split}
\end{equation}
Each null edge $q_i^\mu$ corresponds to a point $P_i$ on $\CI$. In terms of the good cuts $C_{x_i}$, the point associated to $q_i$ is $\{ P_i \} = C_{x_{i+1}} \cap C_{x_i}$. The sequence of momenta $q_1 \to q_2 \to \cdots \to q_n \to q_1$ then defines a sequence of points on $\CI^+$ as $P_1 \to P_2 \to \cdots \to P_n \to P_1$. The boundary quantity that connects these points is the holomorphic frame \eqref{holomorphic_frame}. More precisely, the points $P_i$ and $P_{i+1}$ both live on the good cut $C_{x_{i+1}}$ so the holomorphic frame that parallel transports $P_i \to P_{i+1}$ is $U_{x_{i+1}}(P_{i+1}|P_i)$. Following this observation, it is natural then to conjecture that the boundary dual to the bulk null Wilson loop is
\begin{equation}
\begin{split}
\label{WCn_def}
W[C_n] = \frac{1}{d_R} \Tr_R \left[ U_{x_1}(P_1|P_n) U_{x_n}(P_n|P_{n-1}) \cdots U_{x_2}(P_2|P_1) \right] \,.
\end{split}
\end{equation}
An example of this construction is shown for $n=4$ in Figure \ref{Fig-WLgeom}.
On the boundary, we will compute
\begin{equation}
\begin{split}
W_n = \avg{ W[C_n] } = \frac{1}{Z_{bdy}} \int [ \dt A ] e^{ i S_{bdy} } W[C_n] \,, \qquad Z_{bdy} = \int [ \dt A ] e^{ i S_{bdy} } \,, 
\end{split}
\end{equation}
where $S_{bdy}$ is the boundary Carrollian action. This is evaluated to $\CO(g^2)$ in Section \ref{sec:boundary_calc}.

\subsection{Bulk Calculation}
\label{sec:bulk_calc}

To evaluate the null Wilson loop in the bulk, we use the following trick. Consider the following scaling transformation $\CA \to c \CA$, $g \to c^{-1} g$. Under this scaling transformation,
\begin{equation}
\begin{split}
S_{bulk} \to c^2 S_{bulk} \,, \qquad \CW[C_n] \to \CW[C_n] \,. 
\end{split}
\end{equation}
This implies the identities
\begin{equation}
\begin{split}
\label{identities}
( D_\CA - g \p_g ) S_{bulk} = + 2 S_{bulk}  \,, \qquad ( D_\CA - g \p_g ) \CW[C_n] = 0 \,, \qquad  D_\CA \equiv \int_{M_4} \dt^4 x \CA_\mu^I(x) \frac{\d}{\d \CA_\mu^I(x)} \,.
\end{split}
\end{equation}
Using \eqref{identities}, we find
\begin{equation}
\begin{split}
\label{Wn_diff_eq}
g \p_g \CW_n = - 2 i \int_{M_4} \dt^4 x \avg{ \CL_{bulk}(x) \CW[C_n] }_c \,.
\end{split}
\end{equation}
This relates the expectation value of the Wilson loop to the connected two-point function of the Wilson loop and the Lagrangian operator. Furthermore, while $\CW_n$ is UV divergent (arising from Feynman diagrams where gluon propagators connect to adjacent edges), the two-point function is UV finite and is therefore more straightforward to evaluate. The UV divergences then appear when we integrate over all $x$. In order to avoid discussion of UV divergences, we will evaluate only the integrand in the bulk, and then match it to the analogous quantity on the boundary.

We restrict our calculation to $\CO(g^2)$. To this end, we expand
\begin{equation}
\begin{split}
\CW_n &= 1 + g^2 \CW_n^\2 + O(g^4) \,, \\
\CL_{bulk}(x) &= \CL_2(x) + g \CL_3(x)  + g^2 \CL_4(x)   \,, \\
\CW[C_n] &= 1 -  \frac{g}{d_R} \Tr_R   \oint_{C_n} \CA +  \frac{ g^2 }{ 2d_R} \Tr_R  \CP \oint_{C_n} \CA \oint_{C_n} \CA  + O(g^3)  \,. 
\end{split}
\end{equation}
Substituting these into \eqref{Wn_diff_eq}, we find the equation
\begin{equation}
\begin{split}
\CW_n^\2  =   \int_{M_4} \dt^4 x_0 \CF_n(x_0) \,, \qquad \CF_n \equiv - \frac{i}{2d_R} \Tr_R  \CP \avg{ \CL_2(x_0)  \oint_{C_n} \CA \oint_{C_n} \CA  }_c \,.
\end{split}
\end{equation}
It follows that to $\CO(g^2)$, the Wilson loop integrand only involves the quadratic part of the Lagrangian operator. To $O(g^2)$, the correlator can be evaluated with standard Wick contractions \cite{Alday:2011ga} and the result is
\begin{equation}
\begin{split}
\label{integrandBulk}
\CF_n &= - \frac{iC_R}{4} \CG^{\mu\nu}(x_0) \CG_{\mu\nu}(x_0) \,,  
\end{split}
\end{equation}
where $C_R$ is the quadratic Casimir of the representation $R$ defined by $T_R^I T_R^I = - C_R {\mathbbm 1}$, and $\CG_{\mu\nu}(x_0)$ is defined by
\begin{equation}
\begin{split}
\d^{IJ} \CG_{\mu\nu}(x_0) = \sum_i  \int_0^1 \dt \tau q_i^\rho  \avg{ [ \p_\mu \CA^I_\nu(x_0) - \p_\nu \CA^I_\mu(x_0) ]  \CA^J_\rho(x_i + \tau q_i ) } \,. 
\end{split}
\label{deltaG}
\end{equation}
To evaluate this, we work in $R_\xi$-gauge where the two-point function for the gauge field is
\begin{equation}
\begin{split}
\avg{ \CA_\mu^I(x) \CA_\nu^J(0) } = \frac{\d^{IJ}}{4\pi^2}  \left( \frac{\eta_{\mu\nu}}{x^2 + i \e} - \frac{1}{4} ( 1-\xi  ) \p_\mu \p_\nu \ln \frac{x^2 + i \e}{\ell_0^2} \right)\,,
\end{split}
\end{equation}
where $\ell_0$ is some infrared length-scale. The $\xi$-dependence drops out by symmetry in \eqref{deltaG} and the result is gauge-invariant as expected.
Performing the elementary integral over $\tau$ we find
\begin{equation}
\begin{split}
\CG_{\mu\nu}(x_0) = \frac{1}{2\pi^2} \sum_i ( Y_{i,\mu} Y_{i+1,\nu} - Y_{i,\nu} Y_{i+1,\mu} ) \,, \qquad Y_i^\mu \equiv \frac{x_{i,0}^\mu}{x_{i,0}^2 + i \e } \,, 
\end{split}
\end{equation}
where $x_{ij}^\mu \equiv x_i^\mu - x_j^\mu$. 

To compute $\mathcal{F}_n$, we make use of the relation $x_{ij} \cdot x_{kl} = \frac{1}{2}(x_{il}^2 + x_{jk}^2 - x_{ik}^2 - x_{jl}^2)$. From this we find, after dropping terms which disappear upon telescoping the sums over $i$ and $j$,
\begin{equation}
\begin{split}
\label{FFn}
\CF_n &= - \frac{iC_R}{16\pi^4} \sum_{i<j} \frac{\D_{ij}}{D_{ij} } \,,  
\end{split}
\end{equation}
where
\begin{align}
\label{Dij_def}
\D_{ij} &\equiv x_{i,j}^2 x_{i+1,j+1}^2 - x_{i,j+1}^2 x_{i+1,j}^2 \,, \qquad D_{ij} \equiv ( x_{i,0}^2 + i \e ) ( x^2_{i+1,0} + i \e ) ( x_{j,0}^2 + i \e ) ( x_{j+1,0}^2 + i \e )\,.
\end{align}
Since $\D_{i,i} = \D_{i,i\pm1} = 0$, only terms with $i$ and $j$ non-adjacent contribute in the sum in \eqref{FFn}. 

In Section \ref{sec:boundary_calc}, we will evaluate the same quantity on the boundary and try to match the bulk and boundary results at the integrand level. However, as we have noted in Section \ref{sec:interaction_action}, the boundary interaction action is chiral, so the integrand we will obtain from the boundary calculation will be chiral as well. For the bulk and boundary results to be truly equal, the boundary chiral integrand must differ from the bulk parity-even integrand by a total derivative. Using the fact that the bulk integrand is parity-even and the boundary integrand is chiral, it follows that there is a unique boundary term that relates the two,
\begin{equation}
\begin{split}
\label{bdy_term_addition}
\p_\mu \CC_n^\mu = \frac{C_R}{8} \ve_{\mu\nu\rho\s} \CG^{\mu\nu} \CG^{\rho\s}  \,, 
\end{split}
\end{equation}
with $\ve_{0123}=+1$. Adding this boundary term, we find
\begin{equation}
    \begin{split}
        \CF_n^\text{chiral} = \CF_n + \p_\mu \CC_n^\mu = - \frac{i C_R}{2} \CG^{+\mu\nu} \CG^+_{\mu\nu} \quad \text{where} \quad \CG^+_{\mu\nu} = \frac{1}{2} \left( \CG_{\mu\nu} + \frac{i}{2} \ve_{\mu\nu\rho\s} \CG^{\rho\s} \right) . 
    \end{split}
\end{equation}
Adding the boundary term \eqref{bdy_term_addition} is equivalent to adding a $\t$-term to the Yang-Mills Lagrangian with $\t = \frac{8\pi^2 i}{g^2}$, thereby working with the chiral version of the Yang-Mills Lagrangian,
\begin{equation}
    \begin{split}
        \CL^\text{chiral} = - \frac{1}{2} \Tr [ \CF^+_{\mu\nu} \CF^{+\mu\nu} ] .
    \end{split}
\end{equation}

Since all we have done is add a boundary term, the $O(g^2)$ correction to the Wilson loop remains unchanged
\begin{equation}
    \begin{split}
        \CW_n^\2 =   \int_{M_4} \dt^4 x_0 \CF_n(x_0) = \int_{M_4} \dt^4 x_0 \CF^\text{chiral}_n(x_0) \, .
    \end{split}
\end{equation}
To write $\CF_n^\text{chiral}$ in a more convenient form, it is helpful to introduce a spinorial form of $\CG_{\mu\nu}$,
\begin{equation}
    \CG_{\a \da \b \db} = \CG_{\mu \nu} (\sigma^\mu)_{\a \da} (\sigma^\nu)_{\b \db} =  \e_{\da \db} \CG^+_{\a \b} + \e_{\a \b} \CG^-_{\da \db}\,,
\end{equation}
where $\CG^+_{\a \b} = \CG^+_{\b \a}$ is the self-dual contribution. It is given by
\begin{equation}
    \CG^+_{\a \b}(x_0) = -\frac{1}{4\pi^2} \sum_i \frac{\xi_{i\a} \psi_{i\b} + \xi_{i\b} \psi_{i\a}}{(x_{i,0}^2+i\e)(x_{i+1,0}^2+i\e)}\,,
\end{equation}
where
\begin{equation}
\begin{split}
\label{xi_def}
\psi_{i\a} = - (x_{i,0})_{\a \da}\tilde{\xi}_i^{\da} \,, \qquad    - q_i^\mu ( \s_\mu )_{\a\da} = \xi_{i\a} {\tilde \xi}_{i\da} \,, \qquad {\tilde \xi}_{i\da} = \s_i {\bar \xi}_{i\da} \,, \qquad \s_i \equiv \text{sign}(q_i^0)\,.  
\end{split}
\end{equation}
We then find
\begin{equation}\label{chiralBulk}
\begin{split}
\CG^{+}_{\mu\nu}(x_0) \CG^{+\mu\nu}(x_0) &= \frac{1}{2} \CG^+_{\a \b}(x_0) \CG^{+ \a \b}(x_0) = \frac{1}{4\pi^4} \sum_{i<j} \frac{\avg{ \xi_i \xi_j } \avg{ \psi_i \psi_j }}{ D_{ij}} \,,
\end{split}
\end{equation}
where we have used the Schouten identity $\avg{ \psi_j \xi_i } \avg{ \xi_j \psi_i} =  \avg{ \xi_i \xi_j } \avg{ \psi_i \psi_j } - \avg{ \psi_j \xi_j } \avg{\psi_i \xi_i} $ and we have again dropped terms which telescope in the sum. In summary, we have
\begin{equation}
\begin{split}
\label{Fnchiral_MHV_bulk}
\CF_n^\text{chiral} = - \frac{i C_R}{8\pi^4} \sum_{i<j} \frac{\avg{ \xi_i \xi_j } \avg{ \psi_i \psi_j }}{ D_{ij}} \,.
\end{split}
\end{equation}

\subsection{Boundary Calculation}
\label{sec:boundary_calc}

The conjectured boundary dual to a bulk Wilson loop is the operator \eqref{WCn_def},
\begin{equation}
\begin{split}
W[C_n] = \frac{1}{d_R} \Tr_R \left[ U_{x_1}(P_1|P_n) U_{x_n}(P_n|P_{n-1}) \cdots U_{x_2}(P_2|P_1) \right] \,.
\end{split}
\end{equation}
For the boundary calculation, it will be convenient to work in homogeneous coordinates \eqref{sec:homogeneous_coord}. The homogeneous coordinates of the point $P_i$ are $(\msu_i,\xi_i,{\bar \xi}_i)$ where
\begin{equation}
\begin{split}
\msu_i = \frac{1}{2} \bra{\xi_i} x_i \sket{{\bar \xi}_i} = \frac{1}{2} \bra{\xi_i} x_{i+1} \sket{{\bar \xi}_i}  \,. 
\end{split}
\end{equation}
with $\xi_i$ defined in \eqref{xi_def}. The boundary operator is then
\begin{equation}
W[C_n]= \frac{1}{d_R} \Tr_R \left[ U_{x_1}(\xi_1|\xi_n) U_{x_n}(\xi_n|\xi_{n-1}) \cdots U_{x_2}(\xi_2|\xi_1) \right]\,.
\end{equation}

Now we would like to calculate the expectation value of the operator $W(C_n)$ using the Carrollian action. Thanks to our choice of normalisation we have 
\begin{equation}
\avg{ W[C_n]} = 1 + O(g^2) \,, 
\end{equation}
and here we will restrict our attention to the term of $O(g^2)$. Expanding the holomorphic frames in $W(C_n)$, we need two instances of ${\bar a}$ in order to saturate the two occurrences of $a$ in the interaction term $S_{int}$. These can arise from the same frame $U_{x_i}$  which contributes $I_i$ or from two different frames $U_{x_i}$ and $U_{x_j}$ which contributes $I_{ij}$. The holomorphic frames appearing in $S_{int}$ simply reduce to factors of $1$ at this order. Therefore we have
\begin{equation}\label{Wloop}
\begin{split}
\avg{ W[C_n] } &= 1 + \frac{1}{d_R} \left( - \frac{i}{16\pi^2} \right) \left( - \frac{g}{2\pi i} \right)^2  \int_{M_4} \dt^4 x_0 \biggl\{ \sum_i I_i  + \sum_{i<j} I_{ij} \biggr\} + O(g^4) \\
&= 1  + \frac{i g^2}{64\pi^4 d_R}  \int_{M_4} \dt^4 x_0 \biggl\{ \sum_i I_i  + \sum_{i<j} I_{ij} \biggr\} + O(g^4) \,. 
\end{split}
\end{equation}
The factor of $-\frac{i}{16\pi^2}$ comes from the path integral measure $e^{iS_{int}}$, and the factor of $\left( - \frac{g}{2\pi i} \right)^2$ arises from the two ${\bar a}$ insertions from the frame expansion in the boundary operator \eqref{Ux_holomorphic}. The $O(g^2)$ kernels can be written with the endpoints in either order. Exchanging the endpoints changes the sign of each single-insertion kernel and exchanges the integration variables in the two-insertion kernel. Thus the product of two single-insertion kernels and the integrated two-insertion term are unchanged. Explicitly,
\begin{equation}
\begin{split}
\label{IiIij}
I_i &=  \int_{(C_{x_0})^2}[{\bar \l} \dt {\bar \l}] [{\bar \l}' \dt {\bar \l}'] \avg{ \l \l' }^2  \int_{(C_{x_i})^2} \frac{\avg{ \rho_1 \dt \rho_1 } \avg{ \rho_2 \dt \rho_2 } \avg{ \xi_{i-1} \xi_i }}{\avg{  \xi_{i-1} \rho_1 } \avg{ \rho_1 \rho_2 } \avg{ \rho_2 \xi_i }} G(\l,\l',\rho_1,\rho_2)\,, \\
I_{ij} &= \int_{(C_{x_0})^2}[{\bar \l} \dt {\bar \l}] [{\bar \l}' \dt {\bar \l}'] \avg{ \l \l' }^2 \int_{C_{x_i} \times C_{x_j}} \frac{\avg{ \rho \, \dt \rho } \avg{ \xi_{i-1} \xi_i }}{\avg{  \xi_{i-1} \rho } \avg{ \rho \xi_i }} \frac{\avg{ \rho' \dt \rho' } \avg{ \xi_{j-1} \xi_j }}{\avg{ \xi_{j-1} \rho' } \avg{ \rho' \xi_j }} G(\l,\l',\rho,\rho')  \,,
\end{split}
\end{equation}
and 
\begin{equation}
\begin{split}
\label{G_def}
G(\l,\l',\rho,\rho') = \avg{ \Tr \left[ \frac{\p^2 a(\l)}{\p \msu^2} \frac{\p^2 a(\l')}{\p \msu'^2} \right] \Tr_R \left[ {\bar a}(\rho)  {\bar a}(\rho') \right] } \,.
\end{split}
\end{equation}
As explained in \cite{Opreij:2026bdx}, under the action of a double $\msu$ derivative, the form of the propagator \eqref{A_2pt} implies
\begin{equation}
\begin{split}
\int_{C_{x'}} \frac{\avg{ \rho \dt \rho }}{\avg{ \xi \rho } \avg{ \rho \xi' }} \avg{ \p_\msu^2 a^I(\msu,\l) {\bar a}^J (\msu',\rho) } =   + \frac{i}{\pi} \frac{\avg{ \l \dt \l }}{\avg{ \xi \l } \avg{ \l \xi' }} \frac{\d^{IJ}}{\langle \l |x' - x_0|{\bar \l}]^2}\,.
\end{split}
\end{equation}
Using this, we find
\begin{equation}
\begin{split}
 \label{LCn}
I_i &=  \frac{2 d_R C_R}{\pi^2}   \int_{(C_{x_0})^2} \frac{\avg{ \l \dt \l }  [{\bar \l} \dt {\bar \l}] \avg{ \l' \dt \l' } [{\bar \l}' \dt {\bar \l}'] \avg{ \l \l' }^2 \avg{ \xi_{i-1} \xi_i }}{\avg{ \xi_{i-1} \l} \avg{ \l \l' } \avg{ \l' \xi_i } \langle \l |x_{i,0}| {\bar \l}]^2 \langle \l' |x_{i,0}| {\bar \l}']^2} \,,   \\
 I_{ij}&=  \frac{2 d_R C_R}{\pi^2}  \int_{(C_{x_0})^2}\frac{\avg{ \l \dt \l }  [{\bar \l} \dt {\bar \l}] \avg{ \l' \dt \l' } [{\bar \l}' \dt {\bar \l}'] \avg{ \l \l' }^2 \avg{ \xi_{i-1} \xi_i } \avg{ \xi_{j-1} \xi_j }}{\avg{ \xi_{i-1} \l}  \avg{ \l \xi_i } \avg{ \xi_{j-1} \l' }  \avg{ \l' \xi_j } \langle \l |x_{i,0}| {\bar \l}]^2 \langle \l' |x_{j,0}| {\bar \l}']^2}  \,.  
\end{split}
\end{equation}
where $x_{i,0}^\mu=x_i^\mu - x_0^\mu$. We see that the integrand in \eqref{LCn} has additional singularities when $x_{i,0}$ is null or spacelike. To regulate these singularities, we use the Feynman $i\e$-prescription. As in \cite{Opreij:2026bdx}, to each displacement $x_{i,0}$, we add a small imaginary time component,
\begin{equation}\label{regulator}
\begin{split}
x_{i,0}^\mu \to x^\mu_{i,0,\e} = x^\mu_{i,0} + i \e t^\mu \text{sign} \left( t \cdot x_{i,0} \right) , \qquad t^\mu = (1,0,0,0) \quad \implies \quad  x_{i,0,\e}^2 = x_{i,0}^2 + i \e . 
\end{split}
\end{equation}
This regulates the singularity for generic spacelike separation. For null separation, the prescription describes how we should go across the lightcone.

Focusing on the terms which depend on both $x_i$ and $x_j$, we may perform the $\l'$ integral after first writing
\begin{equation}
\frac{\avg{ \l \l' }^2}{\avg{ \xi_{j-1} \l' } \avg{ \l' \xi_j }} = -\frac{\avg{ \l \xi_{j-1} } \avg{ \l \l' }}{\avg{ \xi_{j-1} \xi_j } \avg{ \l' \xi_{j-1} }} + \frac{\avg{ \l \xi_j } \avg{ \l \l' }}{\avg{ \xi_{j-1} \xi_j } \avg{ \l' \xi_j }}\,.
\end{equation}
The integral over $\l'$ may then be performed using the result \cite{Opreij:2026bdx},
\begin{equation}
\int \frac{\avg{ \rho\, \dt \rho } [\bar{\rho}\, \dt \bar{\rho}]}{\langle \rho | X_\e | \bar{\rho}]^2} \frac{\avg{ \tau \rho }}{\avg{ \kappa \rho }} = -\frac{2 \pi i}{X_\e^2} \frac{\langle \tau |X_\e |\bar{\kappa}]}{\langle \kappa | X_\e | \bar{\kappa}]}\,.
\label{Cintegral}
\end{equation}
Applied to \eqref{LCn}, this yields 
\begin{equation}
\begin{split}
I_i &=  \frac{2d_RC_R}{\pi^2}   \frac{2 \pi i}{x_{i,0,\e}^2} \int_{C_{x_0}} \frac{\avg{ \l \dt \l } [{\bar \l} \dt {\bar \l}]  \avg{ \xi_{i-1} \xi_i }}{\avg{ \xi_{i-1} \l } \langle \l | x_{i,0,\e} | {\bar \l}]^2} \frac{\langle \l | x_{i,0,\e} |\bar{\xi}_i]}{\langle \xi_i | x_{i,0,\e} | \bar{\xi}_i]}\,, \\
I_{ij} &=  \frac{2d_RC_R}{\pi^2}   \frac{(- 2 \pi i)}{x_{j,0,\e}^2}  \int_{C_{x_0}}\frac{\avg{ \l \dt \l } [{\bar \l} \dt {\bar \l}]  \avg{ \xi_{i-1} \xi_i }}{\avg{ \xi_{i-1} \l}  \avg{ \l \xi_i } \langle \l |x_{i,0,\e}| {\bar \l}]^2}    \\
&\qquad \qquad \qquad \qquad \qquad \qquad \times \left(  \frac{\avg{ \l \xi_{j-1} } \langle \l |x_{j,0,\e} | \bar{\xi}_{j-1}]}{\langle \xi_{j-1} | x_{j,0,\e}|\bar{\xi}_{j-1}]} 
- \frac{\avg{ \l \xi_j } \langle \l |x_{j,0,\e} | \bar{\xi}_j]}{\langle \xi_j | x_{j,0,\e}|\bar{\xi}_j]} \right)  \,. 
\end{split}
\end{equation}
We may then perform the integral over $\l$ in exactly the same way. Since $\langle \l|x_{j,0,\e}|\bar\xi_k]$ is linear in $\l$, the Schouten identity gives, for $k=j-1,j$,
\begin{equation}\label{Schouten}
\avg{ \xi_{i-1} \xi_i } \langle \l |x_{j,0,\e}|\bar{\xi}_k] = - \avg{ \xi_i \l } \langle \xi_{i-1} |x_{j,0,\e}|\bar{\xi}_k] - \avg{ \l \xi_{i-1}} \langle \xi_i |x_{j,0,\e}|\bar{\xi}_k]\,,
\end{equation}
and the $\l$ integral for $I_{ij}$ becomes a sum of integrals of the same form as \eqref{Cintegral},
\begin{equation}\label{secondEq}
\begin{split}
\!\! \int_{C_{x_0}} \frac{ \avg{ \l \dt \l } [{\bar \l} \dt {\bar \l}] }{\langle \l |x_{i,0,\e}| {\bar \l}]^2} \biggl\{ 
&\frac{\avg{ \l \xi_{j-1} } \langle \xi_{i-1} |x_{j,0,\e}|\bar{\xi}_{j-1}]}{\avg{ \xi_{i-1} \l}  \langle \xi_{j-1} |x_{j,0,\e}|\bar{\xi}_{j-1}]} 
+
\frac{\avg{ \l \xi_{j-1} } \langle \xi_i |x_{j,0,\e}|\bar{\xi}_{j-1}]}{\avg{ \l \xi_i }  \langle \xi_{j-1} |x_{j,0,\e}|\bar{\xi}_{j-1}]} \\
& \qquad \quad - \frac{\avg{ \l \xi_j } \langle \xi_{i-1} |x_{j,0,\e}|\bar{\xi}_j]}{\avg{ \xi_{i-1} \l}  \langle \xi_j |x_{j,0,\e}|\bar{\xi}_j]} 
-
\frac{\avg{ \l \xi_j } \langle \xi_i |x_{j,0,\e}|\bar{\xi}_j]}{\avg{ \l \xi_i }  \langle \xi_j |x_{j,0,\e}|\bar{\xi}_j]}
\biggr\}\,.
\end{split}
\end{equation}
Combining everything we have
\begin{equation}
\begin{split}
\label{Iij} 
I_{ij} &= -\frac{8d_RC_R}{x_{i,0,\e}^2 x_{j,0,\e}^2}  \biggl\{ - \frac{\langle \xi_{j-1} |x_{i,0,\e}|\bar{\xi}_{i-1}] \langle \xi_{i-1} |x_{j,0,\e}|\bar{\xi}_{j-1}]}{\langle \xi_{i-1} |x_{i,0,\e}|\bar{\xi}_{i-1}] \langle \xi_{j-1} |x_{j,0,\e}|\bar{\xi}_{j-1}]}  + \frac{\langle \xi_{j-1} |x_{i,0,\e}|\bar{\xi}_i] \langle \xi_i |x_{j,0,\e}|\bar{\xi}_{j-1}]}{\langle \xi_i |x_{i,0,\e}|\bar{\xi}_i] \langle \xi_{j-1} |x_{j,0,\e}|\bar{\xi}_{j-1}]} \\
&\qquad \qquad \qquad \qquad \qquad \qquad + \frac{\langle \xi_j |x_{i,0,\e}|\bar{\xi}_{i-1}] \langle \xi_{i-1} |x_{j,0,\e}|\bar{\xi}_j]}{\langle \xi_{i-1} |x_{i,0,\e}|\bar{\xi}_{i-1}] \langle \xi_j |x_{j,0,\e}|\bar{\xi}_j]} - \frac{\langle \xi_j |x_{i,0,\e}|\bar{\xi}_i] \langle \xi_i |x_{j,0,\e}|\bar{\xi}_j]}{\langle \xi_i |x_{i,0,\e}|\bar{\xi}_i] \langle \xi_j |x_{j,0,\e}|\bar{\xi}_j]} \biggr\} \,.
\end{split}
\end{equation}
Note that $I_{ij} = I_{ji}$.

For $I_i$, the analysis is even more straightforward and we obtain
\begin{equation}\label{Ii}
I_i = \frac{8d_RC_R}{x_{i,0,\e}^2} \frac{\avg{ \xi_{i-1} \xi_i } [\bar{\xi}_{i-1} \bar{\xi}_i]}{\langle \xi_i|x_{i,0,\e}|\bar{\xi}_i] \langle \xi_{i-1} |x_{i,0,\e}|\bar{\xi}_{i-1}]}\,.
\end{equation}
\eqref{Iij} and \eqref{Ii} are the final results for our boundary calculation. The boundary integrand that contributes to the Wilson loop operator at $O(g^2)$ is 
\begin{equation}
\begin{split}
\label{Fn_bdy}
F_n(x_0) = \frac{i}{64\pi^4 d_R} \left( \sum_i I_i  + \sum_{i<j} I_{ij} \right) .
\end{split}
\end{equation}

\subsection{Matching the Bulk and Boundary Integrands}
\label{app:analyticmatch}

In this section, we show that\footnote{The derivation presented in this Section was guided by ChatGPT and Claude.}
\begin{equation}
\begin{split}
\label{bulk_bdy_test}
F_n(x_0) = \CF_n^\text{chiral}(x_0)  . 
\end{split}
\end{equation}
where $F_n(x_0)$ is defined in \eqref{Fn_bdy} and $\CF_n^\text{chiral}(x_0)$ is defined in \eqref{Fnchiral_MHV_bulk}. We will start with the boundary integrand and manipulate it until we obtain the bulk integrand. To be precise, we will drop the $\e$-regulator in the formulas and establish \eqref{bulk_bdy_test} as an equality of the rational integrands. We leave a careful analysis of the distributional terms for future work.

Upon dropping the regulator we have e.g. $\psi_{i-1,\alpha}=-(x_{i-1,0})_{\alpha \dot{\alpha}}\tilde{\xi}_{i-1}^{\dot{\alpha}}=-(x_{i,0})_{\alpha \dot{\alpha}}\tilde{\xi}_{i-1}^{\dot{\alpha}}$ so that
\begin{equation}
\label{Iijpsiform}
\begin{split}
&I_{ij} = - \frac{8d_RC_R}{x_{i,0}^2 x_{j,0}^2} \biggl\{ - \frac{\avg{ \xi_{j-1} \psi_{i-1} } \avg{ \xi_{i-1} \psi_{j-1} }}{\avg{ \xi_{i-1} \psi_{i-1} } \avg{ \xi_{j-1} \psi _{j-1} }}
+ \frac{\avg{ \xi_{j-1} \psi_i } \avg{ \xi_i \psi_{j-1} }}{\avg{ \xi_i \psi_i } \avg{ \xi_{j-1} \psi_{j-1} }} \\
&\qquad \qquad \qquad \qquad \qquad \qquad \qquad \qquad \qquad \qquad + \frac{\avg{ \xi_j \psi_{i-1} } \avg{ \xi_{i-1} \psi_j }}{\avg{ \xi_{i-1} \psi_{i-1} } \avg{ \xi_j \psi _j }}
- \frac{\avg{ \xi_j \psi_i } \avg{ \xi_i \psi_j }}{\avg{ \xi_i \psi_i } \avg{ \xi_j \psi_j }} \biggr\} \,,
\end{split}
\end{equation}
and
\begin{equation}\label{Iipsiform}
I_i = \frac{8d_RC_R}{x_{i,0}^2} \frac{\avg{ \xi_{i-1} \xi_i } [\tilde{\xi}_{i-1} \tilde{\xi}_i]}{\avg{ \xi_i\psi_i } \avg{ \xi_{i-1} \psi_{i-1}}}\,.
\end{equation}
For non-adjacent frames, besides the two propagator poles, $I_{ij}$ has four spurious poles at $\langle\xi_a\psi_a\rangle=0$, $a\in\{i-1,i,j-1,j\}$. We will see below how they cancel.

The Schouten identity
$\langle\xi_b\psi_a\rangle\langle\xi_a\psi_b\rangle=\langle\xi_a\psi_a\rangle\langle\xi_b\psi_b\rangle
-\langle\xi_a\xi_b\rangle\langle\psi_a\psi_b\rangle$
turns each ratio in \eqref{Iijpsiform}, with $(a,b)=(i-1,j-1),\,(i,j-1),\,(i-1,j),\,(i,j)$, into $1-B_{ab}$, where
\begin{equation}
B_{ab}\equiv\frac{\langle\xi_a\xi_b\rangle\langle\psi_a\psi_b\rangle}{\langle\xi_a\psi_a\rangle\langle\xi_b\psi_b\rangle}\,,
\qquad B_{ab}=B_{ba}\,,\qquad B_{aa}=0\,.
\end{equation}
The constants cancel among the four terms, leaving a double lattice derivative,
\begin{equation}\label{IijB}
I_{ij}=-\frac{8d_RC_R}{x_{i,0}^2x_{j,0}^2}\,\nabla_i\nabla_jB_{ij}\,,\qquad
\nabla_i\nabla_jB_{ij}\equiv B_{ij}-B_{i-1,j}-B_{i,j-1}+B_{i-1,j-1}\,.
\end{equation}
All spurious poles of $I_{ij}$ are now carried by the denominators of the $B_{ab}$.
Setting $j=i$ in \eqref{IijB} and using $B_{ii}=0$, $B_{ab}=B_{ba}$ and $\langle\psi_{i-1}\psi_i\rangle=x_{i,0}^2\,[\tilde\xi_{i-1}\tilde\xi_i]$
gives
\begin{equation}
I_{ij}\big|_{j=i}=\frac{16d_RC_R}{x_{i,0}^4}\,B_{i-1,i}
=\frac{16d_RC_R}{x_{i,0}^2}\,
\frac{\langle\xi_{i-1}\xi_i\rangle[\tilde\xi_{i-1}\tilde\xi_i]}{\langle\xi_{i-1}\psi_{i-1}\rangle\langle\xi_i\psi_i\rangle}
=2I_i\,,
\end{equation}
so that $\sum_iI_i+\sum_{i<j}I_{ij}=\tfrac12\sum_{i,j}I_{ij}$ and hence
\begin{equation}
\label{Fnablaform}
  F_n(x_0)=\frac{i}{128\pi^4d_R}\sum_{i,j}I_{ij} = -\frac{i C_R}{16 \pi^4} \sum_{i,j} \frac{1}{x_{i,0}^2 x_{j,0}^2} \nabla_i \nabla_j B_{ij}\,. 
\end{equation}

For cyclic sums we have 
\begin{equation}
    \sum_i f_i \nabla_i g_i \equiv \sum_if_i\,(g_i-g_{i-1})=\sum_ig_i\,(f_i-f_{i+1})\,.
\end{equation}
We apply this to (\ref{Fnablaform}) with $f_i=1/x_{i,0}^2$. Using the fact that
$\langle\xi_i\psi_i\rangle=x_{i+1,0}^2-x_{i,0}^2$ we note that 
$f_i-f_{i+1} = \frac{1}{x_{i,0}^2}-\frac{1}{x_{i+1,0}^2}=\frac{\langle\xi_i\psi_i\rangle}{x_{i,0}^2x_{i+1,0}^2}$. We then apply identical reasoning to the sum over $j$ so that the denominators of $B_{ij}$ cancel:
\begin{equation}
\sum_{i,j}I_{ij}=-8d_RC_R\sum_{i,j}B_{ij}\,
\frac{\langle\xi_i\psi_i\rangle\langle\xi_j\psi_j\rangle}{D_{ij}}
=-16d_RC_R\sum_{i<j}\frac{\langle\xi_i\xi_j\rangle\langle\psi_i\psi_j\rangle}{D_{ij}}\,,
\end{equation}
where the summand is symmetric and vanishes for $i=j$. 
This is the step at which the spurious poles disappear. The pole at $\langle\xi_k\psi_k\rangle=0$ is shared by the neighbouring frames $k$ and $k+1$, since $\langle\xi_k|x_{k,0}|\bar\xi_k]=\langle\xi_k|x_{k+1,0}|\bar\xi_k]$, and therefore appears in $I_{kj}$ and $I_{k+1,j}$ for every $j$, including the same-frame terms $I_k$ and $I_{k+1}$. In \eqref{Fnablaform}, the coefficient $B_{kj}$ enters $I_{kj}$ with $\frac{1}{x^2_{k,0}}$ and $I_{k+1,j}$ with $-\frac{1}{x^2_{k+1,0}}$. On the pole, $x^2_{k+1,0}=x^2_{k,0}$, so the residues of the two neighbouring frames are equal and opposite. Their difference supplies exactly the factor $\langle\xi_k\psi_k\rangle$ that removes the pole.
The cancellation leaves only the propagator denominators $D_{ij}$. Hence
\begin{equation}
F_n(x_0)=-\frac{iC_R}{8\pi^4}\sum_{i<j}\frac{\langle\xi_i\xi_j\rangle\langle\psi_i\psi_j\rangle}{D_{ij}}
=\mathcal F_n^{\rm chiral}(x_0)\,.
\end{equation}

\section*{Acknowledgements}

We thank David Skinner and Tim Adamo for useful discussions. JD and PM are supported by the STFC Consolidated Grant ST/C004240/1. BGU is supported by an NSERC grant at the University of Waterloo. This work was supported by the Simons Collaboration on Celestial Holography. The authors acknowledge use of ChatGPT and Claude for verification of several formulas throughout this paper. No scientific ideas, calculations, results, or conclusions were generated by ChatGPT or Claude.

\appendix

\section{The Carrollian Two-Point Function}
\label{app:2pt_gauge}

In this Appendix, we show that the 2-point function that follows from the Carrollian kinetic action, namely
\begin{equation}
\begin{split}
\label{app:kinetic_action}
S_{kin} = - 8i \int_{\CI^+} \dt u \dt^2 z \Tr [ \p_u A_z \CL (A_\bz) ] \,,
\end{split}
\end{equation}
is precisely \eqref{A_2pt}. For this, we first recall the definition of the Hilbert transform,
\begin{equation}
\begin{split}
\CL (A_\bz)(u,z,\bz) \equiv \frac{1}{2\pi} \text{PV} \int_\mrr \dt u' \frac{A_\bz(u',z,\bz)}{u-u'} \,. 
\end{split}
\end{equation}
To determine the two-point function, it is convenient to move to Fourier space,
\begin{equation}
\begin{split}
A_z(u,z,\bz) = \int_\mrr \frac{\dt \o}{2\pi} e^{ i \o u } \a_z(\o,z,\bz) \,, \qquad \a_z(\o,z,\bz) = \int_\mrr \dt u e^{ - i \o u } A_z(u,z,\bz)\,.
\end{split}
\end{equation}
The Hilbert transform in Fourier space is then
\begin{equation}
\begin{split}
\CL (\a_\bz)(\o,z,\bz) &= \int_\mrr \dt u e^{ - i \o u } \CL(A_\bz) (u,z,\bz) \\
&= \int_\mrr \dt u e^{ - i \o u } \frac{1}{2\pi} \text{PV} \int_\mrr \dt u' \frac{A_\bz(u',z,\bz)}{u-u'} \\
&= \int_\mrr \dt u e^{ - i \o u } \frac{1}{2\pi} \text{PV} \int_\mrr \dt u' \frac{1}{u-u'} \int_\mrr \frac{\dt \o'}{2\pi} e^{ i \o' u' } \a_\bz(\o',z,\bz) \\
&= - \frac{i}{2} \text{sign}(\o) \a_\bz(\o,z,\bz) \,.
\end{split}
\end{equation}
Here, we used the fact that
\begin{equation}
\begin{split}
\text{PV}  \int_\mrr \dt u \frac{e^{ - i \o u} }{u}   = - i \pi \text{sign}(\o) \,. 
\end{split}
\end{equation}
We then find that the kinetic action in Fourier space is
\begin{equation}
\begin{split}
S_{kin} = 4 i \int_\mcc \dt^2 z \int_\mrr \frac{\dt \o}{2\pi} |\o| \Tr [ \a_z(\o,z,\bz) \a_\bz(-\o,z,\bz) ] \,.
\end{split}
\end{equation}
Using this and the fact that the path integral weight is $e^{i S_{kin}}$, we find that the Fourier space two-point function is
\begin{equation}
\begin{split}
\avg{ \a^I_z(\o,z,\bz) \a^J_\bw(\o',w,\bw) } = \frac{1}{4|\o|} \d^{IJ} (2\pi) \d(\o+\o') \d^\2(z-w) \,. 
\end{split}
\end{equation}
Finally, we move back to position space and get
\begin{equation}
\begin{split}
\avg{ A^I_z(u,z,\bz) A^J_\bw (u',w,\bw) } &= \int_\mrr \frac{\dt \o}{2\pi} \int_\mrr \frac{\dt \o'}{2\pi} e^{ i \o u +  i \o' u' } \avg{ \a^I_z(\o,z,\bz)  \a^J_\bw (\o',w,\bw) } \\
&=  - \frac{\d^{IJ}}{4\pi} \ln \frac{|u-u'|}{\ell_0} \d^\2(z-w)\,. 
\end{split}
\end{equation}
$\ell_0$ is an infrared cut-off. This is needed since the Fourier transform of $1/|\o|$ is divergent at $\o=0$ and needs to be regulated. The regulated integral we use is
\begin{equation}
\begin{split}
\int_\mrr \frac{\dt \o}{2\pi} \frac{1}{|\o|} \left[ e^{ i \o ( u - u' ) } - e^{ i \o \ell_0} \right] = - \frac{1}{\pi} \ln \frac{|u-u'|}{\ell_0}  \,. 
\end{split}
\end{equation}

\section{Normalisation of the Carrollian Action}
\label{app:action_normalisation}

\subsection{Kinetic Term}
\label{app:kinetic_normalization}

In this section, we derive the normalisation of the kinetic term of the Carrollian action. To do this, we will independently derive the two-point function from the quadratic part of the bulk Yang-Mills Lagrangian, which is
\begin{equation}
\begin{split}
\label{L2_normalization}
\CL_\2 = - \frac{1}{4} ( \p_\mu \CA_\nu^I - \p_\nu \CA_\mu^I )^2 \,. 
\end{split}
\end{equation}
We ignore the $\t$ term as it will not affect the result of this Appendix.

In the free theory, the gauge field $\CA_\mu^I(x)$ admits a mode expansion
\begin{equation}
\begin{split}
\CA_\mu^I(x) = \sum_{h=\pm} \int \frac{\dt^3p}{(2\pi)^3} \frac{1}{2p^0} \left[ \ve_\mu^{(h)}(p)^* \CO^I_{(h)}(p) e^{ i p \cdot x} + \ve_\mu^{(h)}(p) \CO^{I\dagger}_{(h)}(p) e^{ - i p \cdot x}  \right] \,,
\end{split}
\end{equation}
where $p^0 = |\vec{p}|$ and $h=\pm$ are the two helicities. The polarisation vectors are normalised so that
\begin{equation}
\begin{split}
\ve^{(h)}(p) \cdot \ve^{(h')}(p)^*  = \d_{h,h'} \,, \qquad p \cdot \ve^{(h)}(p) = 0 \,. 
\end{split}
\end{equation}
Given our normalisation \eqref{L2_normalization} for the quadratic Lagrangian, the creation and annihilation operators satisfy
\begin{equation}
\begin{split}
\label{commutators}
\left[ \CO^I_{(h)}(p) , \CO_{(h')}^{J\dagger}(p') \right] = \d_{h,h'} \d^{IJ} (2\pi)^3 (2p^0) \d^\3(\vec{p} - \vec{p}\,' ) \,. 
\end{split}
\end{equation}

Using this, we can deduce the gauge field on $\CI^+$. To do this, we parameterise the on-shell massless integration variable $p^\mu$ by $\o q^\mu(w,\bw)$. We take the positive and negative helicity polarisation vectors to be
\begin{equation}
\begin{split}
\label{pol_par}
\ve_\mu^\-(p) \equiv \sqrt{2} \p_\bw q_\mu(w,\bw)\,, \qquad \ve_\mu^\+(p) \equiv \sqrt{2} \p_w q_\mu(w,\bw) \,. 
\end{split}
\end{equation}
These satisfy
\begin{equation}
\begin{split}
\ve^\pmm(p) \cdot \ve^\pmm(p) = 0 \,, \qquad \ve^\+(p)  \cdot \ve^\-(p) = 1 \,, \qquad p \cdot \ve^\pmm(p) = 0 \,. 
\end{split}
\end{equation}
Using this parameterisation, \eqref{commutators} reads
\begin{equation}
\begin{split}
\left[ \CO^I_{(h)}(\o,w,\bw) , \CO_{(h')}^{J\dagger}(\o',w',\bw') \right] = \d_{h,h'} \d^{IJ} (2\pi)^3 \frac{4}{\o} \d(\o-\o')\d^\2(w-w') \,. 
\end{split}
\end{equation}
The Dirac delta function in $z$ is normalised so that
\begin{equation}
\begin{split}
\int_\mcc \dt^2 z \d^\2(z-z') = 1 \,, \qquad \dt^2 z = i \dt z \w \dt \bz \,.
\end{split}
\end{equation}
We can now take $r \to \infty$ using the stationary phase approximation. We then find that the gauge field on $\CI^+$ is
\begin{equation}
\begin{split}
\label{Carrollian_mode_exp}
A_z^I(u,z,\bz) &= - \frac{i}{8\sqrt{2} \pi^2} \int_0^\infty \dt \o  \left[ \CO^I_\+(\o,z,\bz) e^{ - i \o u }  - \CO^{I\dagger}_\-(\o,z,\bz) e^{ i \o u }  \right] \,, \\
A_\bz^I(u,z,\bz) &= - \frac{i}{8\sqrt{2} \pi^2} \int_0^\infty \dt \o  \left[ \CO^I_\-(\o,z,\bz) e^{ - i \o u }  - \CO^{I\dagger}_\+(\o,z,\bz) e^{ i \o u }  \right] \,.
\end{split}
\end{equation}
Using these two results, we find that the non-vanishing vacuum Wightman functions are
\begin{equation}
\begin{split}
\bra{0} A_z^I(u,z,\bz) A_\bw^J(u',w,\bw) \ket{0} &= - \frac{\d^{IJ}}{4\pi} \ln \frac{\e + i ( u - u' ) }{\ell_0} \d^\2(z-w)  \,, \\
\bra{0} A_\bw^J(u',w,\bw) A_z^I(u,z,\bz) \ket{0} &= - \frac{\d^{IJ}}{4 \pi} \ln \frac{\e - i ( u - u' )}{\ell_0}  \d^\2(z-w)\,, 
\end{split}
\end{equation}
where $\e>0$ and $\ell_0$ is an infrared length scale.\footnote{In the computation of the two-point function, we obtain the integral $\int_0^\infty \frac{\dt \o}{\o} e^{ - i \o u }$ which needs to be regulated. We do this by replacing $\int_0^\infty \frac{\dt \o}{\o} e^{ - i \o u } \to \int_{\mu}^\infty \frac{\dt \o}{\o} e^{ - i \o u - \e \o }$ and expanding to leading order in $\mu$. The infrared cutoff $\ell_0$ is $\ell_0 = \mu^{-1} e^{-\g_E}$. $\e$ regulates the UV divergence at $\o=\infty$ and $\ell_0$ regulates the IR divergence at $\o=0$.} The average of the two Wightman functions is \eqref{A_2pt}. 

The time-ordered vacuum correlator is
\begin{equation}
\begin{split}
\bra{0} T \{ A_z^I(u,z,\bz) A_\bw^J(u',w,\bw) \} \ket{0} &= - \frac{\d^{IJ}}{4\pi} \ln \frac{ \e + i |u-u'| }{\ell_0} \d^\2(z-w) \,. 
\end{split}
\end{equation}

\subsection{Interaction Term}
\label{app:interaction_normalization}

We check that the interaction action \eqref{interaction_action} reproduces the bulk tree-level three-point MHV amplitude.\footnote{The $--+$ amplitude vanishes for real on-shell Lorentzian momenta. For this calculation, we work with complexified momenta where $z$ and $\bz$ are independent complex variables. For a non-vanishing $--+$ amplitude, we work in the branch where $\bz_1=\bz_2=\bz_3$.} Using the bulk Yang-Mills action, we find
\begin{equation}
\begin{split}
 \CA_3^{I_1I_2I_3}(--+) = \ve^\-_\mu(q_1) \ve^\-_\nu(q_2) \ve^\+_\rho(q_3) \CV^{I_1I_2I_3;\mu\nu\rho}(q_1,q_2,q_3) ( 2\pi )^4 \d^\4(q_1+q_2+q_3) \,,
\end{split}
\end{equation}
where $\CV^{I_1I_2I_3;\mu\nu\rho}(q_1,q_2,q_3)$ is the 3-point gluon interaction vertex,
\begin{equation}
\begin{split}
\CV^{I_1I_2I_3;\mu\nu\rho}(q_1,q_2,q_3) = g f^{I_1I_2I_3} [ \eta^{\mu\nu} ( q_1 - q_2 )^\rho + \eta^{\nu\rho} ( q_2 - q_3 )^\mu + \eta^{\rho\mu} ( q_3 - q_1 )^\nu ] \,. 
\end{split}
\end{equation}
Parameterising $q_i = \o_i q^\mu(z_i,\bz_i)$ and the corresponding polarisations as \eqref{pol_par}, we find that the amplitude simplifies to
\begin{equation}
\begin{split}
\label{ppm_bulk}
\CA_3^{I_1I_2I_3}(--+) &= \sqrt{2} g f^{I_1I_2I_3} \o_3 z_{12} ( 2\pi )^4 \d^\4(q_1+q_2+q_3) \,. 
\end{split}
\end{equation}

We now compute the same amplitude using the Carrollian action. Inverting \eqref{Carrollian_mode_exp}, we find for $\o>0$ that
\begin{equation}
\begin{split}
\CO^I_\+(\o,z,\bz) &= 4 \sqrt{2} \pi i \int_{-\infty}^\infty \dt u e^{ i \o u} A_z^I(u,z,\bz)  \,, \\
\CO^I_\-(\o, z , \bz) &= 4 \sqrt{2} \pi i \int_{-\infty}^\infty \dt u e^{ i \o u} A_\bz^I(u,z,\bz)   \,. 
\end{split}
\end{equation}
It follows that the MHV amplitude of interest is
\begin{equation}
\begin{split}
 \CA_3^{I_1I_2I_3}(--+) &= \avg{ \CO^{I_1}_\- (\o_1 , z_1 , \bz_1) \CO^{I_2}_\- (\o_2 , z_2 , \bz_2)   \CO^{I_3}_\+ (\o_3 , z_3 , \bz_3) } \\
&= \left( 4 \sqrt{2} \pi i \right)^3 \int_{-\infty}^\infty \dt u_1  \int_{-\infty}^\infty \dt u_2 \int_{-\infty}^\infty \dt u_3 e^{ i ( \o_1 u_1 + \o_2 u_2 + \o_3 u_3 ) } \\
&\qquad \qquad \qquad \qquad \times \avg{ A_{\bz_1}^{I_1} ( u_1 , z_1 , \bz_1) A_{\bz_2}^{I_2} ( u_2 , z_2 , \bz_2) A_{z_3}^{I_3} ( u_3 , z_3 , \bz_3 )  } \,. 
\end{split}
\end{equation}
We evaluate the 3-point function of the gauge field using the Carrollian interaction action \eqref{interaction_action}. To do this, we note that the cubic term in the interaction weight is
\begin{equation}
\begin{split}
e^{ i S_{int} } \big|_{\text{cubic}} &= \frac{ig}{32\pi^3} f^{IJK} \int_{M_4} \!\!\! \dt^4 x \int_{\mcc^3} \dt^2 z \dt^2 z' \dt^2 w    \frac{(z-z')^3}{(z-w)(w-z')} [\p_u^2 A^I_z]_x  [\p_u^2 A^J_{z'}]_x   [ A^K_{\bw} ]_x \,. 
\end{split}
\end{equation}
The three-point function is then obtained by contracting the gauge fields using the two-point function \eqref{A_2pt}. This gives
\begin{equation}
\begin{split}
&\avg{ A_{\bz_1}^{I_1}( u_1 , z_1 , \bz_1) A_{\bz_2}^{I_2} ( u_2 , z_2 , \bz_2) A_{z_3}^{I_3} ( u_3 , z_3 , \bz_3 )  }  \\
&\qquad = - \frac{ig}{1024\pi^6} f^{I_1I_2I_3}  \frac{z_{12}^3}{z_{13} z_{32}} \int_{M_4} \!\!\! \dt^4 x  \frac{\ln \frac{\e + i ( u_3 + x \cdot q(z_3,\bz_3))}{\ell_0}   }{ ( u_1 + x \cdot q (z_1,\bz_1)  - i \e )^2  ( u_2 + x \cdot q ( z_2,\bz_2 ) - i \e )^2 } \,. 
\end{split}
\end{equation}
Here, we have fixed the $i\e$ prescription associated to $\o_i > 0$. It follows that
\begin{equation}
\begin{split}
 \CA_3^{I_1I_2I_3}(--+) &= - \frac{\sqrt{2} g}{(2\pi)^3} f^{I_1I_2I_3}  \frac{z_{12}^3}{z_{13} z_{32}} \int_{-\infty}^\infty \dt u_1  \int_{-\infty}^\infty \dt u_2 \int_{-\infty}^\infty \dt u_3 e^{ i ( \o_1 u_1 + \o_2 u_2 + \o_3 u_3 ) } \\
&\qquad \qquad \qquad \times \int_{M_4} \!\!\! \dt^4 x  \frac{\ln \frac{ \e + i ( u_3 + x \cdot q(z_3,\bz_3) )}{\ell_0} }{ ( u_1 + x \cdot q (z_1,\bz_1) - i \e )^2  ( u_2 + x \cdot q ( z_2,\bz_2 ) - i \e )^2 }  \\
 &= \sqrt{2} g f^{I_1I_2I_3}  \frac{z_{12}^3}{z_{13} z_{32}}  \frac{\o_1 \o_2}{\o_3} (2\pi)^4 \d^\4(q_1+q_2+q_3) \,. 
\end{split}
\end{equation}
We can simplify this further using momentum conservation. For the $(--+)$ amplitude, momentum conservation implies $\bz_1=\bz_2=\bz_3$. The remaining equations for momentum conservation then imply $\o_3 = - ( \o_1 + \o_2 )$ and $z_3 = \frac{\o_1z_1+\o_2z_2}{\o_1+\o_2}$. Using this, we find
\begin{equation}
\begin{split}
\label{ppm_bdy}
 \CA_3^{I_1I_2I_3}(--+)  &= \sqrt{2} g f^{I_1I_2I_3} \o_3 z_{12} (2\pi)^4 \d^\4(q_1+q_2+q_3) \,. 
\end{split}
\end{equation}
This agrees with the bulk result \eqref{ppm_bulk}.

\bibliographystyle{utphys}
\bibliography{RQ-bib}

\end{document}